\documentclass[a4paper,10pt]{article}

\usepackage[pdfborder={0 0 0}]{hyperref}

\newcommand{\Nl}{\mathbb{N}}

\usepackage{amssymb,amsmath,amscd}
\usepackage{color,accents}

\newtheorem{definition}{Definition}[section]
\newtheorem{lemma}[definition]{Lemma}
\newtheorem{proposition}[definition]{Proposition}
\newtheorem{theorem}[definition]{Theorem}
\newtheorem{corollary}[definition]{Corollary}
\newtheorem{remark}[definition]{Remark}

\newenvironment{proof*}{\smallskip\par\noindent\emph{Proof. }
 \ignorespaces}{\hfill$\Box$\smallskip\par\ignorespaces}
\newenvironment{proofsketch*}{\smallskip\par\noindent
 \emph{Sketch of proof. }\ignorespaces}
 {\hfill$\oslash$\smallskip\par\ignorespaces}
\newenvironment{example}{\smallskip\par\noindent
\textbf{Example:\ }}{\hfill$\oslash$\smallskip\par\ignorespaces}

\newcommand{\Rl}{{\mathbb{R}}}
\newcommand{\Cl}{\mathbb{C}}
\newcommand{\Hil}{\mathcal{H}}
\newcommand{\M}{\mathcal{M}}
\newcommand{\K}{\mathcal{K}}

\newcommand{\Hti}{\widetilde{H}}

\newcommand{\Om}{\Omega}

\newcommand{\F}{\mathcal{F}}
\newcommand{\B}{\mathcal{B}}
\newcommand{\tp}[1]{^{\otimes #1}}
\newcommand{\re}{\mathrm{Re}}
\newcommand{\im}{\mathrm{Im}}
\newcommand{\KK}{\mathcal{K}}
\newcommand{\hhat}{\hat{h}}
\newcommand{\hti}{\tilde{h}}
\DeclareMathOperator{\rank}{rank}

\newcommand{\bxi}{{\boldsymbol{\xi}}}
\newcommand{\bmu}{{\boldsymbol{\mu}}}

\newcommand{\ot}{\otimes}

\numberwithin{equation}{section}

\title{Modular nuclearity:\\A generally covariant perspective}

\author{Gandalf Lechner\footnote{School of Mathematics, Cardiff University, Senghennydd Road, Cardiff, CF24 4AG, UK, LechnerG@cardiff.ac.uk}\quad and\quad Ko Sanders\footnote{Institut f\"ur Theoretische Physik, Universit\"at Leipzig, Br\"uderstra{\ss}e 16, 04103 Leipzig, Germany, ko.sanders@itp.uni-leipzig.de}
}

\date{\today}

\begin{document}

\maketitle

\begin{abstract}
A quantum field theory in its algebraic description may admit many irregular states. So far, selection criteria to distinguish physically reasonable states have been restricted to free fields (Hadamard condition) or to flat spacetimes (e.g.\ Buchholz-Wichmann nuclearity). We propose instead to use a modular $\ell^p$-condition, which is an extension of a strengthened modular nuclearity condition to generally covariant theories.

The modular nuclearity condition was previously introduced in Min\-kowski space, where it played an important role in constructive two dimensional algebraic QFT's. We show that our generally covariant extension of this condition makes sense for a vast range of theories, and that it behaves well under causal propagation and taking mixtures. In addition we show that our modular $\ell^p$-condition holds for every quasi-free Hadamard state of a free scalar quantum field (regardless of mass or scalar curvature coupling). However, our condition is not equivalent to the Hadamard condition.
\end{abstract}

\section{Introduction}\label{Sec_Intro}

The observables and the states of a system are the two basic ingredients in any physical theory. In quantum field theory, the observables can conveniently be described as elements of a ${}^*$-algebra, and encode fundamental features such as causality into their algebraic (commutation) relations. The states, essential to make contact with empirical results, are then taken to be expectation value functionals on this algebra of observables.

To ensure a consistent probabilistic interpretation, states are required to satisfy the basic algebraic requirements of linearity, positivity, and normalization. But it is well known that these requirements may admit many states that do not correspond to realistic physical situations, often times because they exhibit too irregular or too singular behavior in their expectation values. The problem of finding criteria to select physically reasonable states, modeling particular situations, or ruling out certain pathologies, therefore has a long history in quantum field theory.

In the case of theories on Minkowski spacetime, one may use the Poincar\'e symmetry to select vacuum states by imposing invariance under this symmetry, and requiring spectral conditions for the energy and momentum operators given by such states \cite{Borchers:1965}. Other well-studied selection criteria are concerned with modeling a localized charge, transforming under a global gauge group \cite{DoplicherHaagRoberts:1969,BuchholzFredenhagen:1982}, or the KMS condition, modeling thermal equilibrium states with respect to some prescribed dynamics \cite{BratteliRobinson:1997,Haag:1996}, in the same way as in statistical mechanics \cite{HaagHugenholtzWinnink:1967}.

\medskip

For quantum field theories on a fixed but curved spacetime manifold \cite{Dimock:1980}, or generally covariant theories, formulated in a consistent manner on a large family of spacetimes \cite{BrunettiFredenhagenVerch:2001,HollandsWald:2015}, the problem of selecting physically reasonable states is even more pronounced: A generic spacetime does not possess any non-trivial symmetries that could serve to select states, or a natural dynamics with respect to which one could ask for equilibrium properties.

In this context, the Hadamard condition on states of free scalar fields \cite{KayWald:1991} is a well-studied criterion. It restricts the short distance singularities of the two-point distribution of the state to match that of the Minkowski vacuum, and thereby encodes a finite renormalized energy density \cite{Wald:1994}. Due to canonical commutation relations, this also restricts the singularities of all higher n-point distributions, allowing the perturbative treatment of interactions \cite{Sanders:2010}. The Hadamard condition can therefore be used to select a physically reasonable class of states in this case. However, this selection relies on the particular structures present in a free field theory, and has no straightforward generalization to more general situations. (See however \cite{Verch:1999} for an attempt.)

To overcome these restrictions, it would be desirable to have a criterion for selecting relevant states, or classes of states, that can be formulated for quantum field theories on general spacetimes, and is not restricted to free field theories and their quasi-free states. In this article, we discuss such a criterion.

\medskip

The main idea is to employ the modular nuclearity condition of Buchholz, D'Antoni and Longo \cite{BuchholzDAntoniLongo:1990} in a suitably generalized manner. This condition originated in the work of Buchholz and Wichmann \cite{BuchholzWichmann:1986}, who imposed a so-called ``energy nuclearity condition'', effectively restricting the number of local degrees of freedom of a Minkowski space quantum field theory. On a mathematical level, this is done by requiring that a certain map, formulated in terms of the Hamiltonian in a vacuum state and a bounded region $O$ in Minkowski spacetime, is nuclear. This criterion was motivated by thermodynamical considerations, and in fact implies reasonable thermal behavior such as the existence of equilibrium states \cite{BuchholzJunglas:1989}.

This condition cannot be used on general curved spacetimes because of the appearance of the Hamiltonian. However, there also exists a local version of it, which instead of the Hamiltonian rather uses modular operators arising from applying Tomita-Takesaki theory \cite{BratteliRobinson:1987} to the local observable algebras. This ``modular nuclearity condition'' \cite{BuchholzDAntoniLongo:1990} can be formulated as follows: Consider an inclusion $\tilde{O}\subset O\subset M$ of bounded open regions in Minkowski spacetime $M$, and a state $\omega$ on a quantum field theory on $M$. Denoting the GNS data by $(\Hil_\omega,\pi_\omega,\Omega_\omega)$, one considers the corresponding inclusion $\M_\omega(\tilde{O})\subset\M_\omega(O)$ of the von Neumann algebras generated by the observables localized in $\tilde{O}$ and $O$, respectively, in the representation $\pi_\omega$. In case $\Omega_\omega$ is cyclic and separating for $\M_\omega(O)$, Tomita-Takesaki modular theory defines the modular operator $\Delta_{O,\omega}$ of $\M_\omega(O)$ w.r.t. the vector~$\Omega_\omega$, which gives rise to some form of ``local dynamics'' on $\M_\omega(O)$~\cite{Borchers:2000}. In this context, the modular nuclearity condition is then the condition that the map
\begin{align}\label{eq:Xi1}
	\Xi:\M_\omega(\tilde{O})\to\Hil_\omega\,,\qquad A\mapsto\Delta_{O,\omega}^{1/4}A\Omega_\omega
\end{align}
is nuclear as a map between Banach spaces, i.e. it can be approximated in norm by a series of rank one operators. This condition is interesting from several points of view -- for example, it implies (for factors) that the inclusion $\M_\omega(\tilde{O})\subset\M_\omega(O)$ is split, which amounts to a strong form of statistical independence of $\M_\omega(\tilde{O})$ and $\M_\omega(O)'$ \cite{FlorigSummers:1997}. Furthermore, it has found application in the construction of models of quantum field theory on two dimensional Minkowski space \cite{BuchholzLechner:2004,Lechner:2008}, and in the analysis of the relation of KMS states at different temperatures \cite{Jkel:2004}.

For the purposes of the present article, it is interesting to note that the modular nuclearity condition is, on the one hand, related to the energy nuclearity condition and thermal behavior for Minkowski space theories \cite{BuchholzDAntoniLongo:1990}, and therefore has a good physical motivation. On the other hand, it has the potential of being applicable to quantum field theories on general curved spacetimes because it does not involve any objects which exist only for Minkowski space theories.

\medskip

In this article, we study a (variant of the) modular nuclearity condition for quantum field theories in curved spacetimes, both in a model-independent setting and also in the context of concrete models. To begin with, we introduce the ``modular $\ell^p$-condition'' in Section~\ref{SSec_Definition} in a form which is suggested by general covariance. It differs from the original modular nuclearity condition insofar as we have to take into account the possibility that our states might not have the Reeh-Schlieder property \cite{Sanders:2009-1}. Furthermore, we ask for stronger nuclearity properties, namely we require that maps like \eqref{eq:Xi1} can be approximated in norm by $n$ dimensional operators with an error that decays faster than any inverse power of $n$. Technically speaking, this amounts to asking that $\Xi$ is an operator of type $\ell^p$ for all $p>0$, see Sect.~\ref{SSec_Definition} for precise definitions. We also replace the
exponent $\frac{1}{4}$ in \eqref{eq:Xi1} by a general value $\alpha\in(0,\frac{1}{2})$.

\medskip

We view the modular $\ell^p$-condition as a condition on a state $\omega$ on a quantum field theory on a general curved spacetime, and proceed to analyze its stability properties in a model-independent setting in Section~\ref{SSec_Stability}. We show that the set of states satisfying the modular $\ell^p$-condition is stable under pullback by morphisms, and under taking finite mixtures. Moreover, we show that the modular $\ell^p$-condition behaves well under spacetime deformation.

The behavior under spacetime deformations has the nice effect that to verify the modular $\ell^p$-condition for a suitable class of states in a generally covariant theory, it suffices to consider particularly simple spacetimes such as ultra-static ones. For such spacetimes, a strong energy nuclearity condition for the theory of a free massive Klein-Gordon field in the GNS representation of its canonical vacuum state was already proven by Verch \cite{Verch:1993_3}.

\medskip

The remaining Sections 3--5 primarily serve to prove that the modular $\ell^p$-condition is satisfied by any quasi-free Hadamard state of the free scalar massive Klein-Gordon field (with or without potential), supporting the claim that this condition can be expected to hold for physically reasonable states. However, along the way we also derive several results that are of independent interest.

\medskip

To begin with, in Section~\ref{Sec_secondq}, we consider an abstract second quantization setting: Starting from a real standard subspace of a complex Hilbert space, and its spatial modular theory, we investigate how modular $\ell^p$ properties lift from the one particle level to the second quantized level. Generalizing results of \cite{BuchholzJacobi:1987} and \cite{Lechner:2005} in the Bose and Fermi case, respectively, we prove that such a lift is possible in both cases. As adequate for a general spacetime without time reflection symmetry, we eliminate certain Minkowski type assumptions from \cite{BuchholzJacobi:1987,Lechner:2005}, and derive new results on stability of inclusions of standard subspaces under conjugations.

Having simplified the modular $\ell^p$-condition to a ``one particle condition'' (in the sense of the one-particle subspace of a Fock representation space), we next investigate the corresponding one-particle problem. Since the ground state of a free scalar field in a standard ultra-static spacetime is intimately related to the geometry of the Cauchy surface $\cal C$, and properties of the modified Laplace-Beltrami operator $A:=-\Delta+m^2$ on $\cal C$, we begin the analysis of the one-particle problem by a detailed study of such operators in Section~\ref{Sec_geo}. In particular, we derive several norm and $\ell^p$ bounds on products of general powers of $A$ and multiplication operators, partially drawing from results of Cheeger, Gromov, and Taylor on finite propagation speed estimates \cite{CheegerGromovTaylor:1982}, and results of Verch on energy nuclearity estimates \cite{Verch:1993_3}.

We then turn our attention to the model of a free Klein-Gordon field in Section~\ref{Sec_freefield}. Here we demonstrate that the modular $\ell^p$-condition holds for every quasi-free Hadamard state (Thm.~\ref{Thm_qfhad}) by expressing the local modular operators in terms of the symplectic form given by the two-point function, and building on the results of the previous sections. We then compare our modular condition with the Hadamard condition and show that there also exist non-Hadamard states that satisfy the modular $\ell^p$-condition in Section~\ref{SSec_Hadamard}.

A discussion of our results in Section~\ref{Sec_Discussion} concludes the article.

\section{The modular $\ell^p$-condition}\label{Sec_Modnuc}

In this section we will introduce the modular $\ell^p$-condition in a generally covariant setting. We will formulate it as a condition on the states of a generally covariant quantum field theory $\mathbf{A}$ \cite{BrunettiFredenhagenVerch:2001}. Recall that such a theory associates a $C^*$-algebra $\mathcal{A}(M)$ to every globally hyperbolic spacetime $M$, and it associates a homomorphism $\mathcal{A}(\psi)$ between $\mathcal{A}(\tilde{M})$ and $\mathcal{A}(M)$ to any embedding $\psi:\tilde{M}\to M$ which preserves the metric and has causally convex range. This association of algebras and morphisms is assumed to be functorial. We do not require that the algebraic homomorphisms are injective \cite{DappiaggiHackSanders:2014}, but they do preserve the unit. As a matter of terminology (cf.\ \cite{FewsterVerch:2012}), we will call a morphism $\psi:\tilde{M}\to M$ an inclusion, when it arises as the canonical inclusion of a subset $\tilde{M}\subset M$ into $M$, and in this case we often write $\mathcal{A}(\tilde{
M})\subset\mathcal{A}(M)$. We call a morphism Cauchy if its range contains a Cauchy surface for $M$, and we call it compact when its range is relatively compact in $M$ and has a non-empty causal complement. The theory is said to satisfy the time-slice axiom if and only if each Cauchy morphism $\psi$ gives rise to an isomorphism $\mathcal{A}(\psi)$.

\bigskip

\subsection{Definition of the modular $\ell^p$-condition}\label{SSec_Definition}

For any state $\omega$ on a $C^*$-algebra $\mathcal{A}$, we can perform the following constructions (see e.g.\ \cite{KadisonRingrose:1986} for a general reference). Let $(\pi_{\omega},\mathcal{H}_{\omega},\Omega_{\omega})$ be the GNS-triple, let $\mathcal{M}_{\omega}:=\pi_{\omega}(\mathcal{A})''$ denote the von Neumann algebra closure of the represented algebra and let $Q_{\omega}$ denote the orthogonal projection onto the subspace
\[
\mathcal{H}'_{\omega}:=\overline{\mathcal{M}'_{\omega}\Omega_{\omega}}.
\]
Then introduce the compression of $\mathcal{M}_{\omega}$ to $\mathcal{H}'_{\omega}$:
\[
\mathcal{N}_{\omega}:=\{Q_{\omega}aQ_{\omega}|\ a\in\mathcal{M}_{\omega}\}.
\]
Note that $Q_{\omega}\in\mathcal{M}_{\omega}$ and that $\mathcal{N}_{\omega}$ is the subalgebra of $\mathcal{M}_{\omega}$ of operators that preserve $\mathcal{H}'_{\omega}$ and that vanish on $(\mathcal{H}'_{\omega})^{\perp}$. Viewed as an operator algebra acting on $\mathcal{H}'_{\omega}$, $\mathcal{N}_{\omega}$ is simply the commutant of $\mathcal{M}_{\omega}'$.

Because $\Omega_{\omega}$ is cyclic for $\mathcal{M}_{\omega}$, it is separating for $\mathcal{M}_{\omega}'$. Viewed as a vector in $\mathcal{H}'_{\omega}$, $\Omega_{\omega}$ is both cyclic and separating for $\mathcal{M}_{\omega}'$ and hence it is also cyclic and separating for $\mathcal{N}_{\omega}$. This allows us to apply the Tomita-Takesaki modular theory and to define the (generally unbounded) Tomita operator on $\mathcal{H}'_{\omega}$ by
\begin{align}\label{eq:SQ}
S_{\omega}Q_{\omega}a\Omega_{\omega}:=Q_{\omega}a^*\Omega_{\omega}
\end{align}
for all $a\in\pi_{\omega}(\mathcal{A})$. We extend this anti-linear operator to $\mathcal{H}_{\omega}$ by setting it to $0$ on $(\mathcal{H}'_{\omega})^{\perp}$, i.e.~$S_{\omega}a\Omega_{\omega}:=Q_{\omega}a^*\Omega_{\omega}$. This operator is closable and we denote the closure by the same symbol. We then have the polar decomposition
\[
S_{\omega}=J_{\omega}\Delta_{\omega}^{\frac12},
\]
where the modular conjugation $J_{\omega}$ is anti-linear, self-adjoint and satisfies $J_{\omega}^2=Q_{\omega}$, whereas the modular operator $\Delta_{\omega}\ge 0$ is positive with kernel $(\mathcal{H}'_{\omega})^{\perp}$. It is therefore uniquely characterized by
\begin{align}\label{eq:Delta-Q}
\|\Delta^{\frac12}_{\omega}a\Omega_{\omega}\|=\|Q_{\omega}a^*\Omega_{\omega}\|
\end{align}
for all $a\in\pi_{\omega}(\mathcal{A})$.

The range of the projection $Q_{\omega}$ always contains $\Omega_{\omega}$. It projects onto the span of $\Omega_{\omega}$ if and only if $\omega$ is a pure state. In general, however, the range of $Q_{\omega}$ can be quite large. When $Q_{\omega}=1$, the von Neumann algebra $\mathcal{M}_{\omega}$ is said to be in a standard representation, which allows the immediate application of the modular theory. For this reason there is often a special interest in states $\omega$ for which this is true, e.g.~restrictions of states with the Reeh-Schlieder property \cite{ReehSchlieder:1961,Sanders:2009-1}. The constructions above, following \cite{Araki:1977}, provide a canonical procedure to obtain a von Neumann algebra $\mathcal{N}_{\omega}$ in a standard representation, thereby bypassing the need to impose additional conditions on the state $\omega$.

To formulate our modular $\ell^p$-condition, we need to introduce some terminology to discuss the approximation of a linear map between two Banach spaces by linear maps of finite rank. For a bounded map $\Xi:\mathcal{B}_1\to \mathcal{B}_2$ between two Banach spaces $\mathcal{B}_i$ one defines the $n$'th approximation number as \cite{Pietsch:1972}
\begin{align}\label{eq:approximation-numbers}
\alpha_n(\Xi):=\inf \{\|\Xi-\Xi_n\||\ \Xi_n\mathrm{\ has\ rank\ } \le n\}.
\end{align}
Given $p>0$, the map $\Xi$ is called $\ell^p$ iff
\begin{align}\label{eq:def-p-quasi-norm}
\|\Xi\|_p:=\left(\sum_{n=0}^{\infty}\alpha_n(\Xi)^p\right)^{\frac{1}{p}}<\infty.
\end{align}
The set of $\ell^p$ maps, denoted $\ell^p(\B_1,\B_2)$ or just $\ell^p$ depending on the desired emphasis, forms a linear space and $\|\cdot\|_p$ is a quasi-norm on it \cite{Pietsch:1972}:
\begin{equation}\label{Eqn_quasinorm}
\|\Xi_1+\Xi_2\|_p\le \max\{2,2^{\frac{2}{p}-1}\}\left(\|\Xi_1\|_p+\|\Xi_2\|_p\right).
\end{equation}
The smaller $p$ is, the better $\Xi$ can be approximated by maps of finite rank, so an $\ell^p$ map is also $\ell^q$ for all $q\ge p$. The composition of an $\ell^p$ map $\Xi$ with a bounded map $B$ is again $\ell^p$, with
\begin{equation}\label{eq:p-ideal}
\|\Xi B\|_p\le \|B\|\ \|\Xi\|_p,\qquad \|B\Xi\|_p\le \|B\|\ \|\Xi\|_p.
\end{equation}
This applies in particular to canonical injections, so we can restrict $\ell^p$ maps to subspaces without increasing the $\ell^p$-quasi-norm. Moreover, the composition of an $\ell^p$ map $\Xi_1$ with an $\ell^q$ map $\Xi_2$ is even $\ell^r$ with $r^{-1}:=p^{-1}+q^{-1}$ and $\|\Xi_1\Xi_2\|_r\le2^{\frac{1}{r}}\|\Xi_1\|_p\|\Xi_2\|_q$. The following lemma is often useful:
\begin{lemma}\label{Lem_lpbound}
If $\Xi_2:\mathcal{B}_1\to \mathcal{B}_2$ and $\Xi_3:\mathcal{B}_1\to \mathcal{B}_3$ are bounded linear maps such that $\|\Xi_3(b)\|\le\|\Xi_2(b)\|$ for all $b\in \mathcal{B}_1$, then $\|\Xi_3\|_p\le\|\Xi_2\|_p$ for all $p>0$.
\end{lemma}
\begin{proof*}
The estimate allows us to define a linear map $B$ from the range of $\Xi_2$ to the range of $\Xi_3$ such that $\Xi_3(b)=B\Xi_2(b)$ and $\|B\|\le 1$. $B$ has an extension to $\mathcal{B}_2$ with $\|B\|\le 1$, by the Hahn-Banach Theorem, and hence $\|\Xi_3\|_p=\|B\Xi_2\|_p\le\|\Xi_2\|_p$.
\end{proof*}

\begin{remark}\label{Rem_lpR}
We will also need to consider real linear $\ell^p$ maps between real Banach spaces, which are defined in a completely analogous way. In this context we denote the $n$'th approximation number by $\alpha^{\mathbb{R}}_n(\Xi)$ and the corresponding quasi-norms $\|\Xi\|_{{\mathbb{R}},p}^p=\sum_{n=0}^\infty \alpha^{\mathbb{R}}_n(\Xi)^p$. The estimates \eqref{Eqn_quasinorm} and \eqref{eq:p-ideal} hold also in this case.

Since every complex linear operator of rank at most $n$ is also a real linear operator of rank at most $2n$, we have $\alpha_{2n}^\Rl(\Xi)\leq \alpha_n(\Xi)$. As the $\alpha_n^{\mathbb{R}}(\Xi)$ decay monotonically in $n$, this implies $\ell^p(\Hil)\subset\ell^p_\Rl(\Hil)$, with $\|\Xi\|_{\Rl,p}\leq 2^{1/p}\|\Xi\|_p$.

For a real linear map $Y:\Hil_1\to\Hil_2$ between real Hilbert spaces, let $\hat{Y}:\hat{\Hil}_1\to\hat{\Hil}_2$ be the complex linear extension to the complexified Hilbert spaces. Note that $\|\hat{Y}\|=\|Y\|$ and that $\hat{Y}$ has complex rank $n$ when $Y$ has real rank $n$. It follows that $\alpha_n(\hat{Y})\le\alpha_n^{\mathbb{R}}(Y)$. Conversely, if a complex linear map $T:\hat{\Hil}_1\to\hat{\Hil}_2$ has complex rank $n$ and $E_2$ is the real-orthogonal projection from $\hat{\Hil}_2$ onto $\Hil_2$, then $T^r:=E_2T|_{\Hil_1}$ is a real linear map of real rank $n$ with $\|Y-T^r\|\le \|\hat{Y}-T\|$ and hence $\alpha_n(\hat{Y})\ge\alpha_n^{\mathbb{R}}(Y)$. It follows that $\|Y\|_{\mathbb{R},p}=\|\hat{Y}\|_p$.

Finally we consider a real linear operator $Y$ on a complex Hilbert space. It can be decomposed into a complex linear and a complex anti-linear part $Y^L=\frac{1}{2}(Y-iYi)$ and $Y^A=\frac{1}{2}(Y+iYi)$, which evidently satisfy $Y^L+Y^A=Y$ and $\|Y^{L/A}\|\leq\|Y\|$. Now let $Y_n$ be approximating real linear operators of $\rank_\Rl Y_n\leq n$. Then $Y_n^L$ is complex linear with $\rank_\Cl Y_n^L\leq n$. But since $\|Y^L-Y_n^L\|\leq\|Y-Y_n\|$, this yields that for $Y\in\ell_\Rl^p(\Hil)$ we also have $Y^L\in\ell^p(\Hil)$, with $\|Y^L\|_p\leq\|Y\|_{\Rl,p}$. If $\Gamma$ is any anti-unitary involution on $\Hil$, we similarly find
$\|Y^A\Gamma-Y_n^A\Gamma\|=\|Y^A-Y_n^A\|\leq\|Y-Y_n\|$ and $\|Y^A\Gamma\|_p\leq\|Y\|_{\Rl,p}$.
\end{remark}

Besides $\ell^p$ maps, it will also be useful to introduce $p$-nuclear maps, which share the feature that they can be well approximated by finite dimensional maps.
\newline\indent
A linear map $\Xi$ between two Banach spaces $\B_1,\B_2$ is called $p$-nuclear if there exist vectors $b_n\in\B_2$ and bounded linear functionals $\varphi_n$ on $\B_1$ such that
\begin{align}\label{eq:nuclear-decomposition}
     \Xi(b)=\sum_{n=1}^\infty \varphi(b)\cdot b_n\,,\qquad b\in\B_1\,,
\end{align}
and
\begin{align}\label{eq:p-nuclearity-index}
     \left(\sum_{n=1}^\infty\|\varphi_n\|^p\|b_n\|^p\right)^{1/p}<\infty\,.
\end{align}
In this case, the infimum of \eqref{eq:p-nuclearity-index} over all decompositions \eqref{eq:nuclear-decomposition} is called the {\em $p$-nuclearity index of $\Xi$}, and denoted $\nu_p(\Xi)$. The set of all $p$-nuclear maps $\B_1\to\B_2$ is denoted $N^p(\B_1,\B_2)$.

This notion of $p$-nuclearity is only meaningful for $0<p\leq1$ \cite{FewsterOjimaPorrmann:2004}.
We next clarify the relation between $\ell^p$ maps and $p$-nuclear maps.

\begin{lemma}\label{lemma:p-nuc-and-ell-p-2}
	Let $\B_1,\B_2$ be Banach spaces, and $0<p\leq1$. Then
	\begin{align}\label{eq:nuclear-inclusion}
		\ell^p(\B_1,\B_2)\subset N^p(\B_1,\B_2)\subset\ell^q(\B_1,\B_2)
	\end{align}
	for any $q>p/(1-p)$, where $\ell^\infty(\B_1,\B_2)$ denotes the compact operators $\B_1\to\B_2$ in the operator norm. There hold the estimates
	\begin{align}\label{eq:nucbound1}
		\nu_p(\Xi)
		&\leq
		c_p\,\|\Xi\|_p\,,\;\;\;\qquad \Xi\in\ell^p(\B_1,\B_2)\,,\\
		\|\Xi\|_q
		&\leq
		c_{p,q}\,
		\nu_p(\Xi)\,,\qquad \Xi\in N^p(\B_1,\B_2)
		\,,
		\label{eq:nucbound2}
	\end{align}
	with the constants $c_p=2^{2+3/p}$ and $(c_{p,q})^q=1+p^q/(1-p)^q\sum_{n=1}^\infty n^{q(1-1/p)}$ with $c_{1,\infty}=1$. In particular,
	\begin{align}\label{eq:s-class}
		\bigcap_{p>0}\ell^p(\B_1,\B_2)=\bigcap_{0<p\leq1}N^p(\B_1,\B_2)\,.
	\end{align}
\end{lemma}
\begin{proof*}
	The first inclusion and the bound \eqref{eq:nucbound1} is proven in \cite[Prop.~8.4.2]{Pietsch:1972} for $0<p\leq1$, see also \cite{FewsterOjimaPorrmann:2004} for the extension to $p>1$.
		\newline\indent
	For the second inclusion, let $\Xi:\B_1\to\B_2$ be $p$-nuclear, $\varepsilon>0$, and pick a $p$-nuclear decomposition $\Xi(b)=\sum_{n=1}^\infty\varphi_n(b)b_n$ with $\sum_{n=1}^\infty\|\varphi_n\|_{\B_1^*}^p\|b_n\|_{\B_2}^p\leq(\nu_p(\Xi)+\varepsilon)^p$. Arranging the $\varphi_n$, $b_n$ in such a way that $n\mapsto r_n:=\|\varphi_n\|_{\B_1^*}\|b_n\|_{\B_2}$ is monotonically decreasing, this gives $r_n^p\leq(\nu_p(\Xi)+\varepsilon)^p/n$.
			\newline\indent
	Now define the map $\Xi_n:\B_1\to\B_2$, $\Xi_n(b):=\sum_{k=1}^{n}\varphi_k(b)b_k$, $n\in\Nl_0$. Then $\Xi_n$ has rank at most $n$, and we can estimate, $n\geq1$, $0<p<1$,
	\begin{align*}
		\alpha_n(\Xi)
		&\leq
		\|\Xi-\Xi_n\|
		\leq\sum_{k=n+1}^\infty r_k
		\leq
		\int_n^\infty dk\,\frac{\nu_p(\Xi)+\varepsilon}{k^{1/p}}
		=
		\frac{p}{1-p}\,(\nu_p(\Xi)+\varepsilon)\,n^{1-1/p}\,.
	\end{align*}
	Taking the limit $\varepsilon\to0$ and raising this expression to a power $q>0$, we have summability over $n\in\Nl$ if $q(1-1/p)<-1$. But this is equivalent to $q>p/(1-p)$, which proves the second inclusion in \eqref{eq:nuclear-inclusion}. Taking into account $\alpha_0(\Xi)=\|\Xi\|\leq\nu_p(\Xi)$, also the bound \eqref{eq:nucbound2} follows.
	
    For $p=1$, the estimate $\alpha_n(\Xi)\leq\sum_{k=n+1}^\infty r_k$ shows that $\alpha_n(\Xi)\to0$ as $n\to\infty$, implying compactness of $\Xi$, and $\|\Xi\|_{\infty}=\|\Xi\|=\alpha_0(\Xi)\le\nu_1(\Xi)$.
	
	The equality \eqref{eq:s-class} now follows from the facts that $\ell^p\subset\ell^q$ for $q>p$, and $p/(1-p)\to0$ as $p\to0$.
\end{proof*}

\medskip

If both $\B_1$ and $\B_2$ are Hilbert spaces, one has $\ell^p(\B_1,\B_2)=N^p(\B_1,\B_2)$ for any $0<p\leq1$. In this case, $\ell^p$ consists of all operators $T$ such that $|T|^p$ is trace class \cite{Pietsch:1972}.

After these mathematical preliminaries, we now come to the definition of the modular $\ell^p$-condition. For any $\alpha\in[0,\frac{1}{2}]$, any inclusion morphism $\psi:\tilde{M}\to M$ and any state $\omega$ on $\mathcal{A}(M)$, we define the linear map
\begin{eqnarray}\label{eq:DefXi}
\Xi^{(\alpha)}_{\tilde{M},M;\omega}:\pi_\omega({\cal A}(\tilde{M}))''\to\mathcal{H}_{\omega}:a\mapsto \Delta_{\omega}^{\alpha}a\Omega_{\omega}.
\end{eqnarray}
The power of $\Delta_{\omega}$ is defined by the spectral calculus on $\mathcal{H}'_{\omega}$, and it is defined to be $0$ on $(\mathcal{H}'_{\omega})^{\perp}$.

\begin{definition}[Modular $\ell^p$-condition]\label{Def_modnuc}
Let $M$ a globally hyperbolic spacetime and $\omega$ a state on $\mathcal{A}(M)$. We say that $\omega$ satisfies the \emph{modular $\ell^p$-condition} if and only if for all $\alpha\in(0,\frac12)$, $p>0$ and all compact inclusions $\iota:O\to M$ and $\tilde{\iota}:\tilde{O}\to O$ the maps $\Xi^{(\alpha)}_{\tilde{O},O;\iota^*\omega}$ are $\ell^p$.
\end{definition}

This condition is a strengthened version of the original modular nuclearity condition of Buchholz, D'Antoni and Longo \cite{BuchholzDAntoniLongo:1990-1}, which asks for $1$-nuclearity of the map $\Xi^{(1/4)}$ in the context of a general inclusion of von Neumann algebras. As we will demonstrate, in the theory of a scalar free field, the maps \eqref{eq:DefXi} are even $\ell^p$ for all $p>0$, for a large class of states. That is, they can be approximated much better by finite dimensional maps, as expressed by the 1-nuclearity condition.
\newline\indent
The modular $\ell^p$-condition may be supplemented by detailed conditions on the behavior of $\|\Xi^{(\alpha)}_{\tilde{O},O;\lambda}\|_p$ in its dependence on $\alpha$, $p$ and the geometry of $\tilde{O}$, $O$ and $M$ (cf. \cite{BuchholzWichmann:1986} for the case of a related energy nuclearity condition). We will not make such additional conditions in this paper, but when possible we will provide estimates on how the $\ell^p$-norms behave under the various operations and constructions that we consider.

\subsection{Stability properties of the modular $\ell^p$-condition}\label{SSec_Stability}

In this section we show that the modular $\ell^p$-condition is stable under certain operations on the states, such as pull-back and taking convex combinations, in a general, model-independent setting. We also demonstrate the good behavior of the modular $\ell^p$-condition under spacetime deformations.

Our first result, stability under pull-back, follows almost immediately from Def.~\ref{Def_modnuc} and general covariance, so we omit its proof:
\begin{lemma}\label{Lem_pullback}
If $\psi:\tilde{M}\to M$ is a morphism and $\omega$ a state on $\mathcal{A}(M)$ which satisfies the modular $\ell^p$-condition, then the pull-back $\tilde{\omega}:=\mathcal{A}(\psi)^*\omega$ also satisfies the modular $\ell^p$-condition, because for any compact inclusions $\iota:O\to \tilde{M}$ and $\tilde{\iota}:\tilde{O}\to O$ we have $\Xi^{(\alpha)}_{\tilde{O},O;\iota^*\tilde{\omega}}=\Xi^{(\alpha)}_{\tilde{O},O;(\psi\circ\iota)^*\omega}$.
\end{lemma}

In order to prove more properties of the modular $\ell^p$-condition we will make use of L\"owner's Theorem for unbounded operators:
\begin{theorem}[L\"owner's Theorem]\label{Thm_Loewner}
Let $I=(a,b)\subset\mathbb{R}$ be an open interval, where $b=\infty$ is allowed, and let $f:I\to\mathbb{R}$ be a continuous function. Then the following two statements are equivalent:
\begin{enumerate}
\item There is a holomorphic function $F$ on the upper half complex plane such that $\mathrm{Im}(F(z))>0$ and which has $f$ as a continuous boundary value on $I$.
\item For all self-adjoint (possibly unbounded) operators $A$, $B$ on a Hilbert space $\mathcal{H}$ with $a<A\le B<b$ (or $a<A\le B$ when $b=\infty$) on the form domain of $B$, we also have $f(A)\le f(B)$ on the intersection of the form domains of $f(A)$ and $f(B)$.
\end{enumerate}
\end{theorem}
When these statements are satisfied, the function $f$ is called operator monotonic.
\begin{proof*}
In the standard version of L\"owner's Theorem one replaces the second statement by a weaker one, where only bounded operators $A$ and $B$ with spectrum in $I$ are allowed \cite{Hansen:2013}. We will not repeat the proof of that result here, but only show that the weaker version of the second statement implies the stronger one. By a translation we may assume that $a=0$, so $0<A\le B$. For $n\in\mathbb{N}$ we set $a_n:=n^{-1}$ and $b_n:=b-n^{-1}$, or $b_n:=n$ when $b=\infty$. We let $E_n$ and $F_n$ be the spectral projections for $A$ and $B$, respectively, onto $[a_n,b_n]$ and we fix $c\in(0,b)$. We then define
\begin{eqnarray}
A_m&:=&E_mAE_m\nonumber\\
A_{n,m}&:=&F_nA_mF_n+c(1-F_n)\nonumber\\
B_n&:=&F_nBF_n+c(1-F_n)\nonumber
\end{eqnarray}
for all $m,n\in\mathbb{N}$ with $m,n> \frac{2}{b}$ when $b$ is finite. We have $\min\{a_n,c\}\le B_n\le\max\{b_n,c\}$, $a_m\le A_m\le b_m$ and hence $\min\{a_m,c\}\le A_{n,m}\le\max\{b_m,c\}$, so all these operators are bounded with spectrum in $I$. From the weak version of the second statement and $A_{n,m}\le F_nAF_n+c(1-F_n)\le B_n$ we then obtain
\[
f(A_{n,m})\le f(B_n)=F_nf(B)F_n+f(c)(1-F_n).
\]
For any fixed $m$ we have $\lim_{n\to\infty}A_{n,m}=A_m$ as a strong limit, and $A_m$ and all $A_{n,m}$ have spectrum in the same compact subset $[\min\{a_m,c\},\max\{b_m,c\}]$ of~$I$. Approximating the continuous function $f$ by polynomials one may therefore show by an $\frac{\epsilon}{3}$-argument that $\lim_{n\to\infty}f(A_{n,m})=f(A_m)=E_mf(A)E_m$ as a strong limit. For every $\psi$ in the form domain of $f(B)$ we then find
\begin{eqnarray}
\langle\psi,E_mf(A)E_m\psi\rangle&=&\lim_{n\to\infty}\langle\psi,f(A_{n,m})\psi\rangle\nonumber\\
&\le&\lim_{n\to\infty}\langle\psi,(F_nf(B)F_n+f(c)(1-F_n))\psi\rangle=\langle\psi,f(B)\psi\rangle.\nonumber
\end{eqnarray}
When $\psi$ is also in the form domain of $f(A)$, then we may take the limit $m\to\infty$ to find the desired equality.
\end{proof*}

\begin{remark}\label{Rem_LoewnerBoundary}
When $f$ is operator monotonic, it is monotonically increasing. If it has a continuous extension to the lower boundary $a$ of the interval $I$ with $f(a)\le0$, then the second statement can be extended to operators $A,B$ such that $a\le A\le B<b$ (or $a\le A\le B$), in which case $f(A)\le f(B)$ on the form domain of $B$. Indeed, the eigenspace of $\sqrt{B+a}$ of eigenvalue $0$ is contained in the eigenspace of $\sqrt{A+a}$ of eigenvalue $0$, so both operators act the same way on this subspace and it remains to consider the orthogonal complement. There, however, we may repeat the proof of Theorem \ref{Thm_Loewner}, supplementing the last line with the remark that $\langle\psi,f(A)\psi\rangle\le \lim_{m\to\infty}\langle\psi,E_mf(A)E_m\psi\rangle$ (because $f(a)\le0$), and $\lim_{m\to\infty}\psi$ is in the form domain of $f(A)$, because $f(A)$ is semi-bounded from below.
\end{remark}

The following corollary involving operators on a subspace is more tailored to our needs:
\begin{corollary}\label{Cor_Loewner}
Let $I=(a,b)$ as in Theorem \ref{Thm_Loewner} and let $f:I\to\mathbb{R}$ be operator monotonic with a continuous extension to $a$ such that $f(a)\le0$. Let $A$ be a self-adjoint operator on a Hilbert space $\mathcal{H}$ and $B$ on a subspace $\mathcal{H}'$. When $a\le A<b$ and $a\le B<b$ (or $a\le A$ and $a\le B$ when $b=\infty$), and when $A\le B$ on the form domain of $B$, then $f(A)\le f(B)$ on the form domain of $f(B)$.
\end{corollary}
\begin{proof*}
We let $P$ denote the orthogonal projection in $\mathcal{H}$ onto $\mathcal{H}'$ and for any $n\in\mathbb{N}$ we let $E_n$ be the spectral projection of $A$ onto $[a,b-n^{-1}]$ (or $[a,a+n]$ when $b=\infty$). We set $A_n:=E_nA+a(1-E_n)$, which is a bounded operator with spectrum in $[a,b)$. For any $\epsilon>0$ we then note that
\begin{eqnarray}
A_n&\le&A_n+(\epsilon P-\epsilon^{-1}(1-P))A_n(\epsilon P-\epsilon^{-1}(1-P))\nonumber\\
&=&(1+\epsilon^2)PA_nP+(1+\epsilon^{-2})(1-P)A_n(1-P).\nonumber
\end{eqnarray}
Because the terms on the right-hand side act on orthogonal subspaces we may apply L\"owner's Theorem \ref{Thm_Loewner} (in the extended version of Remark \ref{Rem_LoewnerBoundary}) to find $Pf(A_n)P\le f((1+\epsilon^2)PA_nP)$. Using the continuity of $f$ and the spectral calculus of the bounded operator $PA_nP$ we may take the limit $\epsilon\to0^+$ to find
\[
Pf(A_n)P\le f(PA_nP).
\]
On the form domain of $B$ we have $PA_nP=A_n\le A\le B$, so by L\"owner's Theorem in the Hilbert space $\mathcal{H}'$ we find $Pf(A_n)P\le f(PA_nP)\le f(B)$ on the form domain of $f(B)$. Taking the limit $n\to\infty$ yields the result.
\end{proof*}

We now apply these results to modular operators to obtain a generalization of \cite[Lemma~2.4]{BuchholzDAntoniLongo:1990-1}:
\begin{lemma}\label{Lem_inclusion}
Let $\mathcal{B}\subset\mathcal{A}$ be an inclusion of $C^*$-algebras and let $\omega$ be a state on $\mathcal{A}$ with restriction $\lambda$ to $\mathcal{B}$. For all $b\in\mathcal{B}$ and $\alpha\in[0,\frac12]$ we then have
\[
\|\Delta_{\omega}^{\alpha}\pi_{\omega}(b)\Omega_{\omega}\|\le\|\Delta_{\lambda}^{\alpha}\pi_{\lambda}(b)\Omega_{\lambda}\|.
\]
\end{lemma}
\begin{proof*}
Let $P$ denote the orthogonal projection in $\mathcal{H}_{\omega}$ onto $\overline{\pi_{\omega}(\mathcal{B})\Omega_{\omega}}$, so we may identify the GNS-representation of $\lambda$ as $\pi_{\lambda}:=P\pi_{\omega}|_{\mathcal{B}}P$, $\mathcal{H}_{\lambda}:=P\mathcal{H}_{\omega}$ and $\Omega_{\lambda}:=\Omega_{\omega}$. Let $Q_{\omega}$ and $Q_{\lambda}$ be the orthogonal projections onto $\mathcal{H}'_{\omega}$ and $\mathcal{H}'_{\lambda}$, where we extend $Q_{\lambda}$ to $\mathcal{H}_{\omega}$ by setting $Q_{\lambda}=Q_{\lambda}P$. Note that $P\in\pi_{\omega}(\mathcal{B})'$ and $\pi_{\lambda}(\mathcal{B})'=(P\pi_{\omega}(\mathcal{B})P)'\supset P\pi_{\omega}(\mathcal{A})'P$. It follows that $\mathcal{H}'_{\lambda}\supset P\mathcal{H}'_{\omega}$ and hence $PQ_{\omega}=Q_{\lambda}PQ_{\omega}=Q_{\lambda}Q_{\omega}$ and $Q_{\omega}P=Q_{\omega}Q_{\lambda}$. Hence,
\[
PQ_{\omega}P=Q_{\lambda}Q_{\omega}Q_{\lambda}\le Q_{\lambda}.
\]
For any $b\in\mathcal{B}$ we have $\pi_{\omega}(b)^*\Omega_{\omega}=P\pi_{\lambda}(b)^*\Omega_{\lambda}$ and therefore, by \eqref{eq:Delta-Q},
\begin{eqnarray}
\langle \pi_{\omega}(b)\Omega_{\omega},\Delta_{\omega}\pi_{\omega}(b)\Omega_{\omega}\rangle&=&
\langle \pi_{\omega}(b)^*\Omega_{\omega},Q_{\omega}\pi_{\omega}(b)^*\Omega_{\omega}\rangle\nonumber\\
&\le&\langle \pi_{\lambda}(b)^*\Omega_{\lambda},Q_{\lambda}\pi_{\lambda}(b)^*\Omega_{\lambda}\rangle\nonumber\\
&=&\langle \pi_{\lambda}(b)\Omega_{\lambda},\Delta_{\lambda}\pi_{\lambda}(b)\Omega_{\lambda}\rangle.\nonumber
\end{eqnarray}
This proves that $\Delta_{\omega}\le\Delta_{\lambda}$ on the form domain of $\Delta_{\lambda}$. The result for $\alpha=0$ or $\alpha=\frac12$ is immediate, and the general result follows from Corollary \ref{Cor_Loewner}, because the function $x^{\beta}=e^{\beta\log(x)}$ has $0^{\beta}=0$ and it is operator monotonic on $x\ge0$ for any $\beta\in(0,1)$ by L\"owner's Theorem.
\end{proof*}

We are now in a position to show that the modular $\ell^p$-condition is preserved under taking convex combinations of states:
\begin{proposition}\label{Prop_convexity}
Let $\omega_1$ and $\omega_2$ be two states on $\mathcal{A}(M)$ and $\omega=r_1\omega_1+r_2\omega_2$ for some $r_1,r_2>0$ with $r_1+r_2=1$. Then $\omega$ satisfies the modular $\ell^p$-condition if both $\omega_i$ do. Moreover, for any two compact inclusions $\iota:O\to M$ and $\tilde{\iota}:\tilde{O}\to O$ we have
\[
\|\Xi^{(\alpha)}_{\tilde{O},O;\iota^*\omega}\|_p\le\max\{2,2^{\frac{2}{p}-1}\}\left(\sqrt{r_1}\|\Xi^{(\alpha)}_{\tilde{O},O;\iota^*\omega_1}\|_p
+\sqrt{r_2}\|\Xi^{(\alpha)}_{\tilde{O},O;\iota^*\omega_2}\|_p\right)
\]
for all $\alpha\in[0,\frac12]$.
\end{proposition}
\begin{proof*}
Denote the pull-backs of the states to $O$ by $\lambda_i:=\iota^*\omega_i$ and note that $\lambda:=\iota^*\omega=r_1\lambda_1+r_2\lambda_2$. Let $\mathcal{H}:=\mathcal{H}_{\lambda_1}\oplus\mathcal{H}_{\lambda_2}$ and $\Omega:=\sqrt{r_1}\Omega_{\lambda_1}\oplus\sqrt{r_2}\Omega_{\lambda_2}$. By construction, the modular operator for $\mathcal{M}:=\pi_{\lambda_1}(\mathcal{A}(O))''\oplus\pi_{\lambda_2}(\mathcal{A}(O))''$ and $\Omega$ is $\Delta:=\Delta_{\lambda_1}\oplus\Delta_{\lambda_2}$. For the map
\[
\Xi^{(\alpha)}:(\pi_{\lambda_1}\oplus\pi_{\lambda_2})({\cal A}(\tilde{O}))''\to\mathcal{H}:a\mapsto\Delta^{\alpha}a\Omega
\]
we then see that $\Xi^{(\alpha)}=\sqrt{r_1}\Xi^{(\alpha)}_{\tilde{O},O;\lambda_1}\oplus\sqrt{r_2}\Xi^{(\alpha)}_{\tilde{O},O;\lambda_2}$. It follows that the left-hand side defines an $\ell^p$ map if and only if both summands on the right-hand side do, and
\begin{eqnarray}
\sqrt{r_i}\|\Xi^{(\alpha)}_{\tilde{O},O;\lambda_i}\|_p&\le&\|\Xi^{(\alpha)}\|_p\nonumber\\
&\le&\max\{2,2^{\frac{2}{p}-1}\}\left(\sqrt{r_1}\|\Xi^{(\alpha)}_{\tilde{O},O;\lambda_1}\|_p+\sqrt{r_2}\|\Xi^{(\alpha)}_{\tilde{O},O;\lambda_2}\|_p\right).\nonumber
\end{eqnarray}

We may identify the GNS-representation of $\lambda$ as $\pi_{\lambda}(a):=\pi_{\lambda_1}(a)\oplus\pi_{\lambda_2}(a)$, with $\Omega_{\lambda}:=\Omega$ and $\mathcal{H}_{\lambda}=\overline{\pi_{\lambda}(\mathcal{A}(\tilde{O}))\Omega}$. Because $\pi_{\lambda}(\mathcal{A}(\tilde{O}))$ is a sub-algebra of $\mathcal{M}$ we may apply Lemma \ref{Lem_inclusion} to see that
\[
\|\Delta_{\lambda}^{\alpha}\pi_{\lambda}(a)\Omega_{\lambda}\|\le\|\Delta^{\alpha}\pi_{\lambda}(a)\Omega_{\lambda}\|.
\]
From Lemma \ref{Lem_lpbound} we then find $\|\Xi^{(\alpha)}_{\tilde{O},O;\lambda}\|_p\le\|\Xi^{(\alpha)}\|_p$, and the conclusion follows.
\end{proof*}

The following corollary tells us that we may always enlarge the larger algebra and/or shrink the smaller algebra without problems. In particular, if $\omega$ satisfies the modular $\ell^p$-condition, then for any compact morphism $\iota:O\to M$ and $\alpha\in(0,\frac12)$ the maps $\Xi^{(\alpha)}_{O,M;\omega}$ are $\ell^p$ for all $p>0$.
\begin{corollary}\label{Cor_inclusions}
Let $\mathcal{B}_1\subset\mathcal{B}_2\subset\mathcal{A}_2\subset\mathcal{A}_1$ be inclusions of $C^*$-algebras and let $\omega_1$ be a state on $\mathcal{A}_1$ with restriction $\omega_2$ to $\mathcal{A}_2$. For all $\alpha\in[0,\frac12]$, $p>0$, the maps
\[
\Xi^{(\alpha)}_i:\mathcal{B}_i\to\mathcal{H}_{\omega_i}:b\mapsto\Delta_{\omega_i}^{\alpha}\pi_{\omega_i}(b)\Omega_{\omega_i}
\]
satisfy $\|\Xi^{(\alpha)}_1\|_p\le\|\Xi^{(\alpha)}_2\|_p$.
\end{corollary}
\begin{proof*}
From Lemma \ref{Lem_inclusion} we have the estimate $\|\Xi^{(\alpha)}_1(b)\|\le \|\Xi^{(\alpha)}_2(b)\|$ for all $b\in\mathcal{B}_1$. Since the injection $\mathcal{B}_1\subset\mathcal{B}_2$ is also bounded with norm $1$, the estimate follows immediately from Lemma \ref{Lem_lpbound}.
\end{proof*}

Corollary \ref{Cor_inclusions} is the algebraic basis of a spacetime deformation argument, which follows a well known pattern
\cite{FullingNarcowichWald:1981,Verch:2001,Sanders:2009-1}. We follow here the recent formulation of \cite{Fewster:2015}, who defines a regular Cauchy pair in a spacetime $M$ to be an ordered pair $(\tilde{V},V)$ of non-empty, relatively compact open subsets of a smooth space-like Cauchy surface $\mathcal{C}$ such that $\overline{\tilde{V}}\subset V$ and $\overline{V}$ has non-empty complement in $\mathcal{C}$. The sets $\tilde{V},V$ define diamond regions $D(\tilde{V}),D(V)$ in $M$, which are globally hyperbolic spacetimes in their own right, and the canonical injections $D(\tilde{V})\subset M$ and $D(V)\subset M$ are morphisms in the category of spacetimes. The following lemma proves the existence of sufficiently many regular Cauchy pairs:
\begin{lemma}\label{Lem_Cauchypairs}
Given any compact inclusions $\iota:O\to M$ and $\tilde{\iota}:\tilde{O}\to O$ there is a regular Cauchy pair $(\tilde{V},V)$ in $M$ such that $\overline{\tilde{O}}\subset D(\tilde{V})$ and $\overline{V}\subset O$.
\end{lemma}
\begin{proof*}
Let $p\in O$ be a point in the causal complement of $\overline{\tilde{O}}$ and let $\mathcal{C}_0$ be a smooth space-like Cauchy surface for $O$ containing $p$. Note that $K:=J(\overline{\tilde{O}})\cap\mathcal{C}_0$ is compact with a non-empty complement in $\mathcal{C}_0$. We may then choose relatively compact open subsets $\tilde{V},V$ in $\mathcal{C}$ such that $K\subset\tilde{V}$, $\overline{\tilde{V}}\subset V$ and $\overline{V}$ has a non-empty complement in $\mathcal{C}_0$, i.e.~$(\tilde{V},V)$ is a regular Cauchy pair in $O$. It follows from Lemma 2.4 in \cite{Fewster:2015} that $(\tilde{V},V)$ is also a regular Cauchy pair in $M$ and the desired inclusions follow from the construction.
\end{proof*}

Our deformation result is then
\begin{theorem}\label{Thm_deformation}
Assume that the theory $\mathbf{A}$ satisfies the time-slice axiom and let $M_1$ and $M_2$ be globally hyperbolic spacetimes with diffeomorphic Cauchy surfaces. Given compact inclusions $\iota:O\to M_1$ and $\tilde{\iota}:\tilde{O}\to O$ and a Cauchy surface $\mathcal{C}_2$ of $M_2$ there is a regular Cauchy pair $(\tilde{V}_2,V_2)$ in $M_2$, contained in $\mathcal{C}_2$, and a chain of Cauchy morphisms
\[
\begin{CD}
M_1 @<\psi_1<< N_1 @>\chi_1>> \tilde{M} @<\chi_2<< N_2 @>\psi_2>> M_2
\end{CD}
\]
such that the isomorphism $\nu:=\mathcal{A}(\psi_2)\mathcal{A}(\chi_2)^{-1}\mathcal{A}(\chi_1)\mathcal{A}(\psi_1)^{-1}$ satisfies
\[
\nu(\mathcal{A}(\tilde{O}))\subset \mathcal{A}(D(\tilde{V}_2))\subset \mathcal{A}(D(V_2))\subset\nu(\mathcal{A}(O)).
\]
It follows that for any $\alpha\in[0,\frac12]$ and any state $\omega_2$ on $M_2$ with $\omega_1:=\nu^*\omega_2$ we have
\[
\|\Xi^{(\alpha)}_{\tilde{O},O;\lambda_1}\|_p\le\|\Xi^{(\alpha)}_{D(\tilde{V}_2),D(V_2);\lambda_2}\|_p,
\]
where $\lambda_1:=\omega_1|_{\mathcal{A}(O)}$ and $\lambda_2:=\omega_2|_{\mathcal{A}(D(V_2))}$.
\end{theorem}
\begin{proof*}
By the time-slice axiom we have $\mathcal{A}(W)=\mathcal{A}(D(W))$ for any causally convex region $W\subset M$. Using Lemma \ref{Lem_Cauchypairs} we find a regular Cauchy pair $(\tilde{V}_1,V_1)$ in $M_1$ such that $\overline{\tilde{O}}\subset D(\tilde{V}_1)$ and $\overline{V_1}\subset O$. Theorem 3.4 in \cite{Fewster:2015} proves the existence of the Cauchy pair $(\tilde{V}_2,V_2)$ and a chain of Cauchy morphisms such that
\[
\nu(\mathcal{A}(D(\tilde{V}_1)))\subset \mathcal{A}(D(\tilde{V}_2))\subset\mathcal{A}(D(V_2))\subset\nu(\mathcal{A}(D(V_1))).
\]
(The Cauchy surface $\mathcal{C}_2$ can be prescribed by Proposition 2.1 of \cite{Fewster:2015}.) Because $\tilde{O}\subset D(\tilde{V}_1)$ and $D(V_1)\subset D(O)$ the first claim follows. The second claim then follows directly from Corollary \ref{Cor_inclusions}.
\end{proof*}
Note that the diamond regions $D(\tilde{V})$ and $D(V)$ themselves may not be relatively compact in $M$, because they may extend too far in the time direction. Nevertheless, one can always cut them down in the time direction to regions $\tilde{W}\subset D(\tilde{V})$ and $W\subset D(V)$ in order to obtain compact inclusions $\tilde{O}\subset \tilde{W}\subset W\subset O$.

\begin{remark}\label{Rem_deformation}
Suppose we are given sets of states $\mathcal{S}_i$ on $\mathcal{A}(M_i)$ such that all states in $\mathcal{S}_2$ satisfy the modular $\ell^p$-condition and all spacetime deformations as in Theorem \ref{Thm_deformation} map all states in $\mathcal{S}_1$ into $\mathcal{S}_2$. Then it is clear from the theorem that all states in $\mathcal{S}_1$ also satisfy the modular $\ell^p$-condition. This argument will be applied in Section \ref{Sec_freefield} to the sets of quasi-free Hadamard states of a free scalar field. It is then sufficient to consider only ultra-static spacetimes $M_2$, because they already cover all possible diffeomorphism classes of Cauchy surfaces.

In fact, the spacetime deformation argument is even stronger than Theorem \ref{Thm_deformation} suggests, because we can also interpolate between free fields with different masses, scalar curvature couplings and other external (non-dynamical) potential energy terms. It then suffices to prove the modular $\ell^p$-condition only for minimally coupled, massive free scalar fields on ultra-static spacetimes, in order to conclude it for any mass, scalar curvature coupling and external potential on any globally hyperbolic spacetime.
\end{remark}


\section{Nuclearity conditions and second quantization}\label{Sec_secondq}

To prepare the investigation of the modular $\ell^p$-condition in free field theories on generic spacetime manifolds, we study in this section $\ell^p$-conditions and nuclearity conditions in an abstract second quantization setting. The main aim is to relate nuclearity properties of the map $\Xi$ \eqref{eq:DefXi}, defined on a second quantization von Neumann algebra, to corresponding ``one-particle conditions''. In the absence of representations of the Poincar\'e group, this term is meant to refer to the single particle subspace of a Fock space which will be introduced subsequently.

Similar questions have been analyzed before, in particular by Buchholz and Wichmann \cite{BuchholzWichmann:1986} for an energy nuclearity condition, and in a more general context, appropriate also for discussing modular nuclearity, by Buchholz and Jacobi~ \cite{BuchholzJacobi:1987}. These authors consider Bosonic systems; the analogous question for the Fermionic case has been settled in \cite{Lechner:2005}. But whereas the setting of these articles is motivated by Minkowski space quantum field theory and ground states, we are interested in fairly general spacetime manifolds and states here. It will therefore be necessary to generalize the known results significantly.

\subsection{$\ell^p$-conditions and Bosonic second  quantization}\label{SSec_secondq}

Let $\F(\Hil)$ denote the Bose Fock space over a complex Hilbert space $\Hil$, with Fock vacuum vector $\Om\in\F(\Hil)$. We denote the projection onto the one-particle space $\Hil\subset\F(\Hil)$ by $P_1$. For $\xi\in\Hil$, we have the usual creation and annihilation operators $a(\xi)^*$, $a(\xi)$ on (a dense domain in) $\F(\Hil)$. Their sum $a(\xi)^*+a(\xi)$ is essentially self-adjoint on this domain, and gives rise to unitary Weyl operators by
\begin{align}\label{eq:Weyl}
	W(\xi)
	=
	\exp\left(i\,(\overline{a(\xi)^*+a(\xi)})\right)\,,
\end{align}
where the bar denotes the self-adjoint closure. The map $\xi\mapsto a(\xi)^*+a(\xi)$ is only real linear, and defines a map from closed real linear subspaces\footnote{In the context of the Klein Gordon quantum field, the subspaces may be related to real Cauchy data with prescribed supports, see Section~\ref{SSec_Hadamard}.} $H\subset\Hil$ to von Neumann algebras $\M(H)\subset\B(\F(\Hil))$ via
\begin{align}\label{eq:second-quantization-vNeumann-algebra}
	\M(H)
	:=
	\{W(h)\,:\, h\in H\}''
	\subset
	\B(\F(\Hil))\,.
\end{align}
We collect some well-known properties of this map in the following lemma. In its formulation, we make use of the symplectic complement $\accentset{\circ}{H}$ of a closed real subspace $H$, taken w.r.t.~the imaginary part of the scalar product of $\Hil$, which is again a closed real linear subspace.

\begin{lemma}\label{Lem_Gamma_delta}
	Let $H\subset\Hil$ be a closed real subspace. Then
	\begin{enumerate}
		\item $\Om$ is cyclic for $\M(H)$ if and only if $H+iH\subset\mathcal{H}$ is dense.
	       \item $\Om$ is separating for $\M(H)$ if and only if $H\cap iH=\{0\}$.
		\item The map $H\mapsto\M(H)$ preserves inclusions.
		\item 
		$\M(H)'=\M({\accentset{\circ}{H}})$.
	\end{enumerate}
\end{lemma}

For more detailed properties of this map, see \cite{Haag:1996}, or \cite[Thm.~I.3.2]{LeylandsRobertsTestard:1978} for a proof.

\medskip

To discuss $\ell^p$- and $p$-nuclearity-properties, we consider in addition to a closed real subspace $H\subset\Hil$ also a self-adjoint (possibly unbounded) linear operator $X$ of second quantized form on $\F(\Hil)$. Later on, $X$ will be taken to be a modular operator of a von Neumann algebra containing $\M(H)$. In the context of energy nuclearity conditions (on ultrastatic spacetimes), one would take $X=e^{-\beta L}$ for some inverse temperature parameter $\beta>0$ and a second quantized Hamiltonian~$L$ \cite{Verch:1993_3}. In the present section, we keep $X$ abstract. When necessary, we will denote its restriction to $\Hil$ by $X_1=X|_\Hil$, and we will assume $H\subset{\rm dom}(X_1)$ and $\M(H)\Om\subset{\rm dom}(X)$. We then define the linear map
\begin{align}\label{eq:Xi-HX}
	\Xi_{H,X_1}:\M(H)\to\F(\Hil)
	\,,\qquad
	A\mapsto XA\Om\,,
\end{align}
similar to the map \eqref{eq:DefXi} appearing in the modular $\ell^p$ condition.

\medskip

Let us next state a theorem of Buchholz and Jacobi \cite[Thm.~2.1]{BuchholzJacobi:1987} about nuclearity properties of $\Xi_{H,X_1}$. Its formulation makes use of {\em conjugations} $\Gamma$ on a complex Hilbert space, which are here defined to be anti-unitary involutions. Given a conjugation $\Gamma$, we write $\Gamma^\pm:=\frac{1}{2}(1\pm \Gamma)$ and note that these are real linear real self-adjoint projections with the obvious properties $\Gamma^++\Gamma^-=1$, $\Gamma^\pm\Gamma^\mp=0$, $\Gamma\Gamma^\pm=\pm\Gamma^\pm$, $i\Gamma^\pm=\Gamma^\mp i$.

\begin{theorem}{\bf\cite{BuchholzJacobi:1987}}\label{theorem:Buchholz-Jacobi}
     Let $H,X$ be as above, satisfying the following two additional assumptions:
     \begin{enumerate}
     	\item There exists a conjugation $\Gamma$ on $\Hil$ which commutes with $X_1$, and two closed complex linear subspaces $\K_\pm\subset\Hil$, such that $\Gamma\K_\pm=\K_\pm$ and
     	\begin{align}\label{eq:Inclusion-Gamma-Structure}
			H&=\Gamma^+\K_++\Gamma^-\K_-\,.			
     	\end{align}
		\item Denoting the (complex linear) projections onto $\K_\pm$ by $E_\pm$, the operators $X_1E_\pm\in\B(\Hil)$ are trace class and satisfy $\|X_1E_\pm\|<1$.
     \end{enumerate}
	Then $\Xi_{H,X_1}$ is nuclear, and its nuclearity index can be estimated as
	\begin{align}\label{eq:buchholz-bound}
		\nu_1(\Xi_{H,X_1})
		\leq
		\det(1-|X_1E_+|)^{-2}\cdot	\det(1-|X_1E_-|)^{-2}
		<
		\infty\,.
	\end{align}
\end{theorem}

Our following generalization of this result involves the real orthogonal projection $E_H$ onto $H$. To define it, we consider $\Hil$ as a real Hilbert space, with scalar product $\re\langle\,\cdot\,,\,\cdot\,\rangle$. This still induces the same norm on $\Hil$, and defines a notion of real adjoint of (real or complex) linear operators on $\Hil$. Since this real adjoint coincides with the usual adjoint for complex linear operators, we will denote it by a superscript $*$ as usual. Then $E_H=E_H^2=E_H^*$ is a real linear (real) self-adjoint projection.

\begin{theorem}\label{theorem:second-quantization}
     In the notations above, the following hold true:
     \begin{enumerate}
     	\item If $X_1E_H$ is $\ell^p_\Rl(\Hil)$ for some $0<p\leq1$, and $\|X_1E_H\|<1$, then $\Xi_{H,X_1}$ is $p$-nuclear and $\ell^q$ for $q>p/(1-p)$. In particular, if the assumption holds for all $p>0$, then $\Xi_{H,X_1}$ is $\ell^q$ for all $q>0$.
     	\item $\|X_1E_H\|_{\Rl,p}\le\sqrt{e}\,2^{1/p}\,\|\Xi_{H,X_1}\|_p$ for all $p>0$.
     	\end{enumerate}
\end{theorem}

There are three differences between Thm.~\ref{theorem:second-quantization} and Thm.~\ref{theorem:Buchholz-Jacobi}. First, the assumption $a)$ of Thm.~\ref{theorem:Buchholz-Jacobi} is absent in Thm.~\ref{theorem:second-quantization}. Second, different spectral density conditions ($p$-nuclearity and $\ell^p$, for complex respectively real linear operators) are used. Third, we also demonstrate the necessity of one of our assumptions in part $b)$. We did not try to derive a sharp bound on the $p$-nuclearity index or $\ell^q$-quasi-norms of $\Xi_{H,X_1}$. However, from the proof given later, one sees that in the situation of Thm.~\ref{theorem:second-quantization}, one has a bound of the form
\begin{align}
	\nu_p(\Xi_{H,X_1})^p
	\leq
	\prod_{l=1}^\infty t_l^{-p}\,{\rm Li}_{-p}(t_l^p)
	<\infty\,,
\end{align}
where Li is the polylogarithm and the $t_l$ are the eigenvalues of a positive operator $T\in\ell^p$ constructed from $X_1E_H$, satisfying $\|T\|<1$ and $\|T\|_p\leq c_p\|X_1E_H\|_{\Rl,p}$ for some numerical constant $c_p$, cf.~\eqref{eq:xibound}.

\medskip

We begin with a discussion of assumption $a)$ of Thm.~\ref{theorem:Buchholz-Jacobi}. To this end, it is useful to characterize inclusions of the form \eqref{eq:Inclusion-Gamma-Structure} in a more invariant manner.

\begin{lemma}\label{lemma:Hstructure}
	\begin{enumerate}
		\item Let $\Gamma$ be a conjugation on $\Hil$. Then a closed real subspace $H\subset\Hil$ is of the form $H=\Gamma^+\K_++\Gamma^-\K_-$ with two closed complex linear subspaces $\K_\pm\subset\Hil$ which are invariant under $\Gamma$ if and only if $\Gamma H=H$.
		\item If $H$ is a closed real subspace as in $a)$, the real orthogonal projection $E_H$ onto $H$ is related to the complex orthogonal projections $E_\pm$ onto $\K_\pm$ by
		\begin{align}
			E_H=\Gamma^+ E_++\Gamma^- E_-\,.
		\end{align}
	\end{enumerate}
\end{lemma}
\begin{proof*}
	$a)$ 	Suppose $H$ has the form $H=\Gamma^+\K_++\Gamma^-\K_-$ as described. Then, as $\Gamma^2=1$, it follows immediately that $\Gamma H=H$.
	
	For the other implication, assume that $\Gamma H=H$, and define $\K_\pm:=\Gamma^\pm H+i\,\Gamma^\pm H$. These are two complex linear subspaces which are both invariant under $\Gamma$, and we claim that they are also closed. In fact, if $\xi_n:=\Gamma^\pm h_n+i\Gamma^\pm\hhat_n$ is a Cauchy sequence in $\K_\pm$, then so is
     \begin{align*}
	  \Gamma^\pm\xi_n
	  =
	  \Gamma^\pm\Gamma^\pm h_n+\Gamma^\pm i\,\Gamma^\pm\hhat_n
	  =
	  \Gamma^\pm h_n\,.
     \end{align*}
	Hence $\Gamma^\pm h_n$ and $\Gamma^\pm\hhat_n$ are Cauchy sequences, and the closedness of $H$ implies the closedness of $\K_\pm$.

     Using the same properties of $\Gamma$ again, we also see that
     \begin{align*}
	  \Gamma^+\K_++\Gamma^-\K_-
	  &=
	  \Gamma^+(\Gamma^+H+i\Gamma^+H)+\Gamma^-(\Gamma^-+i\Gamma^-H)
	  \\
	  &=
	  \Gamma^+H+\Gamma^-H\,.
     \end{align*}
     But as $\Gamma^++\Gamma^-=1$ and $\Gamma H=H$, we have $\Gamma^+H+\Gamma^-H=H$, i.e. $H$ is of the claimed form $H=\Gamma^+\KK_++\Gamma^-\KK_-$.

     $b)$ Since the $\K_\pm$ are invariant under $\Gamma$, this conjugation commutes with the projections $E_\pm$. Using this fact, it is straightforward to check that $Q:=\Gamma^+E_++\Gamma^-E_-$ is a self-adjoint real linear projection. In view of $H=\Gamma^+\K_++\Gamma^-\K_-$, this space is pointwise invariant under $Q$. On the other hand, if $Q\xi=\xi$ for some $\xi\in\Hil$, then $\Gamma^\pm \xi=\Gamma^\pm E_\pm\xi$. Thus $\xi=\Gamma^+\xi+\Gamma^-\xi\in\Gamma^+\K_++\Gamma^-\K_-=H$. This implies that $Q$ and $E_H$ coincide.
%
%
\end{proof*}
\medskip

The situation described in part $a)$ of this lemma is generic: As we will show later, any closed real subspace $H$ admits a conjugation $\Gamma$ such that $\Gamma H=H$ (Prop.~\ref{proposition:existence-of-gamma}). Furthermore, by virtue of the spectral theorem in its multiplication operator form \cite[Thm.~VIII.4]{ReedSimon:1972}, any (complex linear) self-adjoint operator $X_1$ is unitarily equivalent to an operator multiplying with a real-valued function on some $L^2$-space. Considering pointwise complex conjugation on that space, it follows that there exists a conjugation $\Gamma$ commuting with $X_1$.

But in general, there does not exist a conjugation commuting with $X_1$ and preserving $H$ at the same time, as it is assumed in Thm.~\ref{theorem:Buchholz-Jacobi}. We will show later in Section~\ref{SSec_secondqmod} that such a conjugation does also not always exist if $X_1$ is taken to be the modular operator suggested from the modular $\ell^p$-condition. This complication of a missing suitable conjugation can be circumvented in our proof of Thm.~\ref{theorem:second-quantization} below, but results in less stringent bounds on the $\ell^p$ quasi norms.

Before we can proceed to the proof of Thm.~\ref{theorem:second-quantization}, we need a technical lemma.
\bigskip
\begin{lemma}\label{lemma:T-Gamma}
	Let $T_\pm$ be two bounded complex linear operators, and $\Gamma$ a conjugation that commutes with both of them. Define the real linear operator
	\begin{align}
		T:=\Gamma^+T_++\Gamma^-T_-
		\,.
	\end{align}
	\begin{enumerate}
		\item There holds the norm equality
		\begin{align}\label{eq:T-Norms}
			\|T\|=\max\{\|T_+\|,\|T_-\|\}\,.
		\end{align}
		\item Let $p>0$. Then $T\in\ell^p_\Rl(\Hil)$ if and only if $T_\pm\in\ell^p(\Hil)$, and in this case, the corresponding $\ell^p$-quasi-norms satisfy the bounds
		\begin{align}\label{eq:p-bound-real-vs-complex}
			\|T\|_{\Rl,p}
			&\leq
			c_p\,\left(\|T_+\|_p+\|T_-\|_p\right)
			\\
			\|T_\pm\|_p
			&\leq
			c_p'\,\|T\|_{\Rl,p}
			\,,\label{eq:p-bound-complex-vs-real}
		\end{align}
		where $c_p,c_p'$ are numerical constants depending only on $p$.
	\end{enumerate}
\end{lemma}
\begin{proof*}
	$a)$ The proof of \eqref{eq:T-Norms} is based on the fact that for a conjugation $\Gamma$ on $\Hil$ and two arbitrary vectors $\psi,\xi\in\Hil$, there always holds
	\begin{align}\label{eq:Gamma-Pythagoros}
		 \|\Gamma^+\xi+\Gamma^-\psi\|^2
		=
		\|\Gamma^+\xi\|^2+\|\Gamma^-\psi\|^2\,,
	\end{align}
	because $\Gamma^\pm$ are real orthogonal projections with $\Gamma^\pm\Gamma^\mp=0$.

	To begin with, note that it readily follows from our assumptions that $T\Gamma^\pm=T_\pm\Gamma^\pm$, and in particular $\|T_\pm\Gamma^\pm\|\leq\|T\|$. But by complex linearity of $T_\pm$, we also have $-iTi=T_+\Gamma^-+T_-\Gamma^+$ and hence $T_\pm\Gamma^\mp=(-iTi)\Gamma^\mp$. This implies $\|T_\pm\Gamma^\mp\|\leq\|T\|$. Using these bounds and \eqref{eq:Gamma-Pythagoros}, we obtain, $\xi\in\Hil$,
	\begin{align*}
		\|T_\pm\xi\|^2
		&=
		\|\Gamma^+T_\pm\Gamma^+\xi+\Gamma^-T_\pm\Gamma^-\xi\|^2
		\\
		&\leq
		\|T_\pm\Gamma^+\|^2\|\Gamma^+\xi\|^2+\|T_\pm\Gamma^-\|^2\|\Gamma^-\xi\|^2
		\\
		&\leq
		\|T\|^2(\|\Gamma^+\xi\|^2+\|\Gamma^-\xi\|^2)
		\\
		&=
		\|T\|^2\cdot\|\xi\|^2\,.
	\end{align*}
	Hence $\|T_\pm\|\leq\|T\|$. On the other hand,
	\begin{align*}
		\|T\xi\|^2
		&=
		\|\Gamma^+T_+\Gamma^+\xi+\Gamma^-T_-\Gamma^-\xi\|^2
		\\
		&=
		\|T_+\Gamma^+\xi\|^2+\|T_-\Gamma^-\xi\|^2
		\\
		&\leq
		\max\{\|T_+\|^2,\|T_-\|^2\}\cdot(\|\Gamma^+\xi\|^2+\|\Gamma^-\xi\|^2)
		\\
		&=
		\max\{\|T_+\|^2,\|T_-\|^2\}\cdot\|\xi\|^2\,,
	\end{align*}
	which implies $\|T\|\leq\max\{\|T_+\|,\|T_-\|\}$. Together with $\|T_\pm\|\leq\|T\|$, this yields \eqref{eq:T-Norms}.
	
	$b)$ From Remark \ref{Rem_lpR} we see that $T_\pm\in\ell^p(\Hil)$ implies $T\in\ell^p_\Rl(\Hil)$, with $\|T\|_{\Rl,p}\leq 2^{1/p}\,c_p(\|T_+\|_p+\|T_-\|_p)$. After renaming $c_p$, this shows \eqref{eq:p-bound-real-vs-complex}. Furthermore, for $T=\Gamma^+T_++\Gamma^-T_-\in\ell_\Rl^p$ a quick calculation shows that in this case, $T_\pm=T^L\pm T^A\Gamma$. Hence
	\begin{align*}
		\|T_\pm\|_p
		&=
		\|T^L\pm T^A\Gamma\|_p
		\leq
		c_p\,\left(	\|T^L\|_p+
		\|T^A\Gamma\|_p
		\right)
		\leq
		2\,c_p\,\|T\|_{\Rl,p}\,,
	\end{align*}
	which completes the proof of \eqref{eq:p-bound-complex-vs-real}.
\end{proof*}
\medskip
Now we are ready for the proof of the main result of this section, Thm.~\ref{theorem:second-quantization}.
\begin{proof*}
$a)$ As explained above, we first need to account for the possibility that there is no conjugation $\Gamma$ such that $[\Gamma,X_1]=0$ and $\Gamma H=H$. We therefore start with a construction to introduce some additional structure. Let $\Gamma$ be a conjugation on $\Hil$, and consider
\begin{align}\label{eq:doubledsystem}
	\underline{\Hil}:=\Hil\oplus{\Hil}
	\,,\qquad
	\underline{\Gamma}
	:=
	\left(
		\begin{array}{cc}
			0&\Gamma\\ \Gamma&0
		\end{array}
	\right)
	\,,\qquad
	\underline{H}:=H\oplus \Gamma H\,,\qquad
    \underline{X}_1:=X_1\oplus \Gamma X_1\Gamma\,.
\end{align}
It is clear that $\underline{\Gamma}$ is a conjugation on $\underline{\Hil}$. Moreover, $\underline{\Gamma}$ leaves the closed real subspace $\underline{H}$ invariant and commutes with $\underline{X}_1$. The real linear projection onto $\underline{H}$ is $E_{\underline{H}}=E_H\oplus \Gamma E_H\Gamma$, which implies $\|\underline{X}_1E_{\underline{H}}\|=\|X_1E_H\oplus \Gamma(X_1E_H)\Gamma\|<1$ by our norm assumption on $X_1E_H$, and
$\|\underline{X}_1E_{\underline{H}}\|_{\Rl,p}\le \max\{4,2^{\frac{2}{p}}\}\|X_1E_H\|_{\Rl,p}$ by the quasi-norm property (\ref{Eqn_quasinorm}).

We now use the natural unitary map implementing the equivalence $\F({\underline{\Hil}})\cong\F(\Hil)\otimes\F({\Hil})$, which carries the Fock vacuum $\underline{\Om}$ of $\F(\underline{\Hil})$ onto $\Om\ot\Om$, and the von Neumann algebra $\M(\underline{H})$ onto $\M(H)\ot \M(\Gamma H)$.

Under this identification, we have $$\Xi_{\underline{H},\underline{X}_1}=\Xi_{H,X_1}\otimes\Xi_{\Gamma H,\Gamma X_1\Gamma }\,.$$ But clearly the maps $F:\M(H)\to\M(H)\ot\M(\Gamma H)$, $A\mapsto A\ot1$ and $G:\F(\Hil)\ot\Om\to\F(\Hil)$, $\Psi\ot\Om\mapsto\Psi$, are linear and bounded, with norm one, and $\Xi_{H,X_1}=G\,\Xi_{\underline{H},\underline{X}_1}F$. Hence $\nu_p(\Xi_{H,X_1})\le\nu_p(\Xi_{\underline{H},\underline{X}_1})$ and
$\|\Xi_{H,X_1}\|_q=\|G\,\Xi_{\underline{H},\underline{X}_1}F\|_q\leq\|\Xi_{\underline{H},\underline{X}_1}\|_q$. It now suffices to prove the claim for the underlined objects.

Lemma~\ref{lemma:Hstructure} applies to $\underline{\Gamma}$, $\underline{H}$, so that we may write $\underline{H}=\underline{\Gamma}^+\K_++\underline{\Gamma}^-\K_-$ and $E_{\underline{H}}=\underline{\Gamma}^+E_++\underline{\Gamma}^-E_-$ with complex linear subspaces $\K_\pm\subset\underline{\Hil}$ and corresponding complex linear projections $E_\pm$, commuting with $\underline{\Gamma}$. Thus $T:=\underline{X}_1E_{\underline{H}}$ has the form assumed in Lemma~\ref{lemma:T-Gamma}, with $T_\pm=\underline{X}_1E_\pm$, and we conclude
\begin{align}
	\|\underline{X}_1E_\pm\|\leq\|\underline{X}_1E_{\underline{H}}\|<1
	\,,\qquad 	\left\|\underline{X}_1E_\pm\right\|_p
	\leq
	c_p'\cdot	\left\|\underline{X}_1E_{\underline{H}}\right\|_{\Rl,p}<\infty\,,
\end{align}
with some numerical constant $c_p'<\infty$.

For $p=1$, the space $\ell^p(\underline{\Hil})$ coincides with the trace class on $\underline{\Hil}$. In that situation, all assumptions of Thm.~\ref{theorem:Buchholz-Jacobi} are satisfied, and we can immediately conclude that $\Xi_{H,X_1}$ is (1-)nuclear, with the bound
\begin{align}\label{eq:buchholz-bound}
	\nu_1(\Xi_{\underline{H},\underline{X}_1})\leq
	\det(1-|\underline{X}_1E_+|)^{-2}\cdot\det(1-|\underline{X}_1E_-|)^{-2}
	<\infty\,.
\end{align}

For general $p$, we need to re-examine the argument underlying Thm~\ref{theorem:Buchholz-Jacobi}. One step in that proof is the construction of a certain joint least upper bound of $\underline{X}_1E_{\pm}$ \cite[p.~316-317]{BuchholzJacobi:1987}. Going through the construction, it becomes apparent that it works for $\ell^p$-operators as well: If $\underline{X}_1E_{\pm}\in\ell^p(\Hil)$, then there exists a positive operator $T\in\ell^p(\Hil)$ such that $$\|T\|\leq\max\{\|\underline{X}_1E_+\|,\|\underline{X}_1E_-\|\}<1$$ and $T^2\geq|\underline{X}_1E_{\pm}|^2$.

To estimate the approximation numbers of $\Xi_{\underline{H},\underline{X}_1}$, we can then follow the argument in \cite{BuchholzWichmann:1986}: Let $\{\xi_k\}_k$ denote an orthonormal basis of $\underline{\Hil}$ consisting of eigenvectors of $T$, i.e.~$T\xi_k=t_k\cdot\xi_k$, with $\sum_kt_k^p<\infty$. Let $\{\bxi_\bmu\}_\bmu$ denote the corresponding ``occupation number'' orthonormal basis of $\F(\underline{\Hil})$, i.e.~$\bmu:\Nl\to\Nl_0$ are summable functions. Then
\begin{align*}
	\sigma_\bmu:=
	\sup_{A\in\M(\underline{H})\setminus\{0\}}
	\frac{|\langle\bxi_\bmu,\,\Xi_{\underline{H},\underline{X}_1}(A)\rangle|}{\|A\|}
	\leq
	\prod_{l=1}^{\infty}(\mu_l+1)t_l^{\mu(l)}
	\,,
\end{align*}
cf.~\cite[p.~338]{BuchholzWichmann:1986}. This implies
\begin{align}\label{eq:polylog}
	\sum_\bmu \sigma_\bmu^p
	\leq
	\sum_\bmu\prod_{l=1}^{\infty}(\mu_l+1)^pt_l^{p\mu(l)}
	=
	\prod_{l=1}^\infty\sum_{m=0}^\infty
	(m+1)^pt_l^{pm}
	=
	\prod_{l=1}^\infty t_l^{-p}\text{Li}_{-p}(t_l^p)
	\,,
\end{align}
where Li denotes the polylogarithm. To show that this expression is finite, it is sufficient to estimate $t_l^{-p}\text{Li}_{-p}(t_l^p)$ for large enough $l$. Recall that $(m+1)^{1/m}\leq e$ for all $m\in\Nl_0$, and thus $(m+1)^pt_l^{pm}\leq (e\,t_l)^{mp}$. Since $t_l\to 0$ monotonically as $l\to\infty$, we have $e\,t_l<1$ for $l$ larger than some $L\in\Nl$. Hence, for large enough~$l$, we have $\sum_{m=0}^\infty(m+1)^pt_l^{mp}\leq(1-(e\,t_l)^p)^{-1}$. As $(e\,t_l)^p$ is summable in $l$, this shows that the product \eqref{eq:polylog} converges. Note that for $p=1$, \eqref{eq:polylog} reduces to the familiar expression $\prod_{l=1}^\infty(1-t_l)^{-2}$ underlying \eqref{eq:buchholz-bound}.

We have therefore found a $p$-nuclear decomposition \eqref{eq:nuclear-decomposition} of $\Xi_{\underline{H},\underline{X}_1}$, and conclude that this map is $p$-nuclear, with $p$-nuclearity index bounded by
\begin{align}\label{eq:xibound}
	\nu_p(\Xi_{\underline{H},\underline{X}_1})^p
	\leq
	\prod_{l=1}^\infty t_l^{-p}\text{Li}_{-p}(t_l^p)
	<\infty\,.
\end{align}
Whereas up to this point, the value of $p>0$ was arbitrary, we now restrict to the case $0<p\leq1$ to apply Lemma~\ref{lemma:p-nuc-and-ell-p-2}, which then tells us that $\Xi_{H,X_1}$ is also $\ell^q$ for any $q>p/(1-p)$.

We remark that in the situation at hand, one can exploit the particular form of our $p$-nuclear decomposition in terms of an orthonormal basis of a Hilbert space to show that $\Xi_{\underline{H},\underline{X}_1}$ is even $\ell^q$ for any $q>2p/(2-p)$.
\bigskip

$b)$ We now prove the second statement, so we may assume that $\Xi_{H,X_1}$ is $\ell^p$ for some $p>0$ (otherwise the estimate is trivially true). We use the fact that a map $\Xi:\mathcal{B}_1\to\mathcal{B}_2$ has the same operator norm and rank as its dual $\Xi^*:\mathcal{B}_2^*\to\mathcal{B}_1^*$, so if $\Xi$ is $\ell^p$ for some $p>0$, then so is $\Xi^*$. Combining this with Lemma \ref{Lem_lpbound} we see that $\|(P_1\Xi_{H,X_1})^*\|_{\Rl,p}\le\|\Xi_{H,X_1}\|_{\Rl,p}$.

Now let $\chi\in\Hil$ be in the domain of $X$. Assume for the moment that $f:=E_HX_1\chi$ is non-zero and define $r:=\|f\|^{-1}$. Writing $\chi^*:=\langle\chi,.\rangle\in\mathcal{H}^*$ we may use $A:=W(rf)\in\mathcal{M}(H)$ with $\|A\|=1$ to estimate
\begin{eqnarray}
\|(P_1\Xi_{H,X_1})^*\chi^*\|&\ge&|((P_1\Xi_{H,X_1})^*\chi^*)(A)|\nonumber\\
&=&|\chi^*(P_1\Xi_{H,X_1}A)|=|\langle\chi,P_1XA\Omega\rangle|\nonumber\\
&=&e^{-\frac12}|\langle\chi,X_1rf\rangle|=e^{-\frac12}r|\langle X_1\chi,E_H^2f\rangle|\nonumber\\
&\ge&e^{-\frac12}r|\re\langle X_1\chi,E_H^2X_1\chi\rangle|\nonumber\\
&=&e^{-\frac12}r\|E_HX_1\chi\|^2=e^{-\frac12}\|E_HX_1\chi\|,\nonumber
\end{eqnarray}
where we used the fact that the projection $E_H$ is real self-adjoint. It follows from this estimate that $\|(X_1E_H)^*\chi^*\|=\|E_HX_1\chi\|\le \sqrt{e}\|(P_1\Xi_{H,X_1})^*\chi^*\|$. The same estimate holds when $f=0$, so it holds on a dense domain in $\Hil$. Hence, $\|(X_1E_H)^*\|\le \sqrt{e}\|(P_1\Xi_{H,X_1})^*\|$, which means that $(X_1E_H)^*$ is bounded and
\begin{align*}
	\|X_1E_H\|_{\Rl,p}
	&=
	\|(X_1E_H)^*\|_{\Rl,p}\le\sqrt{e}\|(P_1\Xi_{H,X_1})^*\|_{\Rl,p}=\sqrt{e}\|P_1\Xi_{H,X_1}\|_{\Rl,p}
	\\
	&\le
	\sqrt{e}\|\Xi_{H,X_1}\|_{\Rl,p}
\end{align*}
by Lemma \ref{Lem_lpbound}. Using the same argument as in Lemma~\ref{lemma:T-Gamma}, we also find $\|\Xi_{H,X_1}\|_{\Rl,p}\leq2^{1/p}\|\Xi_{H,X_1}\|_p$, from which the result follows.
\end{proof*}

\subsection{Second quantization of modular operators}\label{SSec_secondqmod}

We now wish to apply the results of the previous subsection to the case where $X$ is the modular operator of a second quantized von Neumann algebra, containing a subalgebra corresponding to the real subspace considered so far.

As before, we consider a closed real subspace $H$ of a complex Hilbert space~$\Hil$, and denote the symplectic complement of $H$ by $\accentset{\circ}{H}$. Furthermore, we will need to work with two different orthogonal complements, a real and a complex one. The {\em complex}  orthogonal complement of $H$ refers to the scalar product of $\Hil$. It is denoted $H^\perp$, and seen to coincide with $H^\perp=\accentset{\circ}{H}\cap i\accentset{\circ}{H}$ by an elementary calculation. The {\em real} orthogonal complement of $H$, referring to the real scalar product $\re\langle\,\cdot\,,\,\cdot\,\rangle$, was introduced before. We will write the real orthogonal complement of $H$ as $H^{\perp_\Rl}$, and note that $H^{\perp_\Rl}=i{\accentset{\circ}{H}}$.

\medskip

The natural setting of spatial modular theory is that of {\em standard subspaces} (see \cite{Longo:2008} for an overview). A closed real subspace $H\subset\Hil$ is called standard if
\begin{align}
	\overline{H+iH}&=\Hil\,,\label{eq:Hcyclic}
	\\
	H\cap iH&=\{0\}\,.\label{eq:Hseparating}
\end{align}
Thanks to these properties, any standard subspace $H$ has a well-defined densely defined Tomita operator~$S_H$,
\begin{align}\label{eq:def-sh}
	S_H:H+iH=:D_H\to D_H\,,\qquad h_1+ih_2\mapsto h_1-ih_2\,.
\end{align}
As usual, the polar decomposition of this anti-linear involution will be denoted $S_H=J_H\Delta_H^{1/2}$, with $J_H$ an anti-unitary involution, and $\Delta_H$ a complex linear positive operator, satisfying $J_H\Delta_H J_H=\Delta_H^{-1}$. As all our standard subspaces will be only real-linear, we drop the term ``real'' and refer to them simply as standard subspaces.

The second quantized von Neumann algebra $\M(H)$ of a standard subspace~$H$ has the Fock vacuum $\Om$ as a cyclic separating vector (Lemma~\ref{Lem_Gamma_delta}). The modular data of $(\M(H),\Om)$ are closely related to the modular data of $H$ \cite{EckmannOsterwalder:1973}:
\begin{lemma}\label{Lem_Gamma_delta-ii}
	Let $H\subset\Hil$ be a standard subspace. Then
	the modular data $J,\Delta$ of $(\M(H),\Om)$ are related to $J_H,\Delta_H$ by second quantization:
		\begin{align}
			J=\bigoplus_{n=0}^\infty J_H\tp{n}
			\,,\qquad
			\Delta=\bigoplus_{n=0}^\infty \Delta_H\tp{n}\,.
		\end{align}
\end{lemma}

For an inclusion of standard subspaces $\Hti\subset H\subset\Hil$, this shows that taking $X_1=\Delta_H^\alpha$, $0<\alpha<\frac{1}{2}$, and the subspace $\Hti$, we are in the situation described in Thm.~\ref{theorem:second-quantization} for discussing nuclearity properties of $\Xi_{\Hti,\Delta_H^\alpha}$ \eqref{eq:Xi-HX}.

In line with the situation described in Section~\ref{SSec_Definition}, we will however need to consider more general closed real subspaces $H$, which do not necessarily satisfy \eqref{eq:Hcyclic} or \eqref{eq:Hseparating}. In that case, $H$ can be compressed to a standard subspace, as we describe now.

Note that $H^\perp=\accentset{\circ}{H}\cap i\accentset{\circ}{H}$ and $\accentset{\circ}{H}^\perp=H\cap iH$ are closed complex subspaces that are orthogonal to each other. Hence there exists an orthogonal (complex linear) projection $R_H$ such that
\begin{align}\label{eq:H-split}
	\Hil = H^\perp \oplus \accentset{\circ}{H}^\perp \oplus R_H\Hil\,.
\end{align}
In this decomposition, $H=\{0\}\oplus \accentset{\circ}{H}^\perp \oplus R_HH$, i.e. $R_HH\subset H$ is the (complex) orthogonal complement of $H\cap iH$ in $H$, and therefore separating. Considered as a subspace of $R_H\Hil$, the projected real space $R_HH$ is therefore standard~\cite{LeylandsRobertsTestard:1978}. Analogously to Section~\ref{SSec_Definition}, we now define the Tomita operator $S_H$ of a general closed real subspace by
\begin{align}\label{eq:generalized-tomita}
	S_H := 0\oplus0\oplus S_{R_HH}\,,
\end{align}
referring to the decomposition \eqref{eq:H-split}. 

We are now in the position to apply Thm.~\ref{theorem:second-quantization} to the modular setting.

\begin{theorem}\label{thm:modular-nuclearity}
	Let $\Hti\subset H\subset\Hil$ be an inclusion of closed real subspaces, and $0<\alpha<\frac{1}{2}$. If $\|\Delta^\alpha_HE_{\Hti}\|<1$ and $\Delta^\alpha_HE_{\Hti}$ is $\ell_\Rl^p(\Hil)$ for all $p>0$, then $\Xi_{\Hti,\Delta_H^\alpha}$ \eqref{eq:Xi-HX} is $\ell^p$ for all $p>0$.
\end{theorem}
\begin{proof*}
	In view of the split \eqref{eq:H-split}, the Bose Fock space over $\Hil$ has the form
	\begin{align}
		\F(\Hil)
		=
		\F(H^\perp)\otimes\F(\accentset{\circ}{H}^\perp)\otimes\F(R_H\Hil)\,,
	\end{align}
	with Fock vacuum $\Om=\Om^\perp\otimes\accentset{\circ}{\Om}^\perp\otimes\Om_H$ in an obvious notation.
	
	Furthermore, in this decomposition, the second quantized von Neumann algebra $\M(H)$ and its commutant are \cite{LeylandsRobertsTestard:1978}
	\begin{align}
		\M(H)\label{eq:MH}
		&=
		\Cl\otimes\B(\F(\accentset{\circ}{H}^\perp))\otimes\M(R_HH)
		\,,\\
		\M(H)'
		&=
		\B(\F(H^\perp))\otimes\Cl\otimes\M(R_HH)'
		\,.\label{eq:MH'}
	\end{align}
	According to the definitions of Section~\ref{SSec_Definition}, the modular data of $(\M(H),\Om)$ are constructed by first projecting to the subspace generated by $\M(H)$, which is $\Om^\perp\otimes\F(\accentset{\circ}{H}^\perp)\otimes\F(R_H\Hil)$. Then, in this subspace, we consider the projection onto the subspace generated by the commutant $\M(H)'$, which is $\Om^\perp\ot\accentset{\circ}{\Om}^\perp\ot\F(R_H\Hil)$. But on the last tensor factor, $\M(R_HH)$ is based on a standard subspace, with modular operator $\Delta$ the second quantization of $\Delta_{R_HH}$.
	
	This implies that the modular operator of $\M(H)$ w.r.t. $\Om$ is the second quantization of $\Delta_H=S_H^*S_H$, and thus given by $|\Om^\perp\rangle\langle\Om^\perp|\ot|\accentset{\circ}{\Om}^\perp\rangle\langle\accentset{\circ}{\Om}^\perp|\ot\Delta$.
	From this we see that the map $\Xi_{\Hti,\Delta_H^\alpha}$ has the form
	\begin{align*}
		\Xi_{\Hti,\Delta_H^\alpha}:\Cl\otimes\B(\F(\accentset{\circ}{H}^\perp))\otimes\M(R_HH)&\to\F(H^\perp)\otimes\F(\accentset{\circ}{H}^\perp)\otimes\F(R_H\Hil)\,,
		\\
		\lambda\otimes B\otimes A&\mapsto \lambda\Om^\perp\ot\langle\accentset{\circ}{\Om}^\perp,B\accentset{\circ}{\Om}^\perp\rangle \accentset{\circ}{\Om}^\perp\otimes\Delta^\alpha A\Om_H\,.
	\end{align*}
	Under the assumptions made, we know by Thm.~\ref{theorem:second-quantization} that the map $A\mapsto\Delta_H^\alpha A\Om_H$, acting on the rightmost factor, is $\ell^p$ for all $p>0$. Since the other two factor maps are of rank one, the claim follows.
\end{proof*}

\medskip


We wish to address two more topics: The norm bound appearing in the assumptions of Thm.~\ref{theorem:second-quantization}, and the existence of a conjugation commuting with $\Delta_H^\alpha$ and preserving $\Hti$ (cf.~discussion after Lemma~\ref{lemma:Hstructure}).

\medskip

The norm bound required in Thm.~\ref{theorem:second-quantization}~$a)$ is almost a consequence of the $\ell^p$-properties. As we will see in the applications to quantum field theory models, the appearing standard subspaces are typically ``factors'' in the sense that $H\cap{\accentset{\circ}{H}}=\{0\}$, in which case the norm bound is a consequence.

\begin{lemma}\label{Lem_strictnorm}
Let $\Hti\subset H$ be an inclusion of closed real subspaces and $0<\alpha<\frac{1}{2}$. Then $\|\Delta_H^\alpha E_{\Hti}\|\le1$, and $|\Delta_H^\alpha E_{\Hti}|<1$ whenever $\Hti\cap{\accentset{\circ}{H}}=\{0\}$. If in addition $\Delta_H^\alpha E_{\Hti}$ is compact (or even $\ell^p$ for some $p>0$), then
\begin{align}
 \|\Delta_H^\alpha E_{\Hti}\|<1\,.
\end{align}
\end{lemma}
\begin{proof*}
Let $\chi,\eta\in\Hil$ and $h:=E_H\eta$. As $H={\rm dom}\Delta_H^{1/2}$, it follows that the function $f(z):=\langle \chi,\Delta_H^{-iz}h\rangle$ is analytic on the strip $0<\im(z)<\frac{1}{2}$ and continuous on the closure of this strip. In view of $|f(z)|\leq\|\chi\|\|\Delta_H^{\im(z)}h\|$ we see furthermore that $f$ is bounded. Moreover, on the boundary we have $|f(t)|\leq\|\chi\|\|h\|$, and $|f(t+\frac{i}{2})|\leq\|\chi\|\|\Delta_H^{1/2}h\|=\|\chi\|\|J_Hh\|=\|\chi\|\|h\|$, $t\in\Rl$. Hence we may apply the three lines theorem \cite[Thm.~3.7]{Conway:1978} to the effect that $|f(z)|\leq\|\chi\|\|h\|$ throughout the closed strip. Since $\chi$ and $\eta$ are arbitrary, this entails $\|\Delta_H^{-iz}E_H\eta\|\le\|E_H\eta\|\le\|\eta\|$ and hence $\|\Delta_H^{\alpha}E_{\Hti}\|\le\|\Delta_H^{\alpha}E_H\|\le1$.

Now suppose that $\hti\in \Hti$ satisfies $\|\Delta_H^{\alpha}\hti\|=\|\hti\|$. Using the orthogonal decomposition $\hti=R_H\hti+\hti'$ with $R_H\hti\in R_HH\subset H$ and $\hti'\in H$ we then find from the first paragraph that $\|\hti\|=\|\Delta_H^{\alpha}R_H\hti\|\le\|R_H\hti\|$ and hence $\hti'=0$ and $\hti\in \Hti\cap R_H\Hti$. Then note that for $0<\alpha<\frac12$ the estimate $|f(t+i\alpha)|<\|\chi\|\|\Delta_H^{\alpha}\hti\|$ is strict, unless $f$ is constant (see, for example, \cite[Cor.~3.9]{Conway:1978}). Taking $\chi=\hti$, $f$ is the expectation value of a unitary one parameter group, which is constant if and only if $\Rl\ni t\mapsto\Delta_H^{-it}\hti$ is constant. But that would imply $\hti\in{\accentset{\circ}{R_HH}}$ \cite[Prop.~2.1.14]{Longo:2008}, and thus $\hti\in \Hti\cap R_H\Hti\cap{\accentset{\circ}{R_HH}}=\Hti\cap{\accentset{\circ}{H}}$. When this is assumed to be $\{0\}$, we find $|\Delta_H^\alpha E_{\Hti}|<1$, i.e.~$|\Delta_H^\alpha E_{\Hti}|$ has no eigenvectors with eigenvalue 1.

Now suppose that the real linear operator $T:=\Delta_H^\alpha E_{\Hti}$ is compact. Similar to the complex linear case, a real linear compact operator can be represented as $T=\sum_n t_n \re\langle\varphi_n,\,\cdot\,\rangle\cdot\psi_n$ with two real orthonormal bases $\{\varphi_n\}_n$ and $\{\psi_n\}_n$ w.r.t. $\re\langle\,\cdot\,,\,\cdot\,\rangle$, and positive numbers $t_n$ such that $\|T\|=t_1\geq t_n$ for all $n$, and $t_n\to0$ as $n\to\infty$. (Such a representation of $T$ can be established by considering the complexification $\widehat{\Hil}\supset\Hil$ of $\Hil$. Then $T$ gives rise to a complex linear compact operator $\widehat{T}$ on $\widehat{\Hil}$, which leaves $\Hil$ invariant. The claimed representation of $T$ then follows from the canonical form of complex linear compact operators, and restriction to $\Hil\subset\widehat{\Hil}$.)
	
To show $\|T\|<1$, it is therefore sufficient to show that any eigenvalue of the real linear operator $T^*T=\sum_n t_n^2\re\langle\psi_n,\,\cdot\,\rangle\cdot\psi_n$ is strictly less than 1, which follows from the first two paragraphs.
\end{proof*}

\medskip

We now come to our discussion of conjugations. In the proof of Thm.~\ref{theorem:second-quantization} we constructed a doubled Hilbert space with some complex conjugation $\Gamma$. It may happen that such a conjugation already exists without doubling the Hilbert space. In this case our estimates on the $\ell^p$ quasi norms can be improved significantly, essentially by taking a square root.

In Thm.~\ref{theorem:Buchholz-Jacobi}, the existence of a conjugation $\Gamma$, commuting with $X_1$ and preserving the real subspace $H$, is assumed. This has its motivation in the theory of the Klein-Gordon field on Minkowski space in its vacuum representation. When formulated in terms of its time zero field and momentum, consider the von Neumann algebra $\cal N$ generated by the time zero fields, with arbitrary support on the time zero surface $\cal C$. This is a maximally abelian second quantized von Neumann algebra, i.e. ${\cal N}=\M(H_0)=\M({\accentset{\circ}{H}_0})$ for some standard subspace $H_0={\accentset{\circ}{H}_0}$ in the single particle space. As a consequence, the modular operator $\Delta_{H_0}=1$ is trivial, and hence $S_{H_0}=J_{H_0}$ is a conjugation, corresponding to complex conjugation for functions on $\cal C$. This conjugation preserves the time zero fields, smeared with real test functions in a given region $O\subset\cal C$, but changes the sign of the time zero momenta, smeared with
real test functions with support in $O$. Therefore the standard subspace $H(O)$ corresponding to $O''$ has
the structure assumed in Thm.~\ref{theorem:Buchholz-Jacobi}.

\medskip

In general, one can show given a closed real linear subspace $H\subset\Hil$, there always exists a conjugation preserving it. Note that this statement is non-trivial because $H$ is only real linear.

\begin{proposition}\label{proposition:existence-of-gamma}
	Let $H\subset\Hil$ be a closed real subspace. Then there exists a conjugation $\Gamma$ on~$\Hil$ such that $\Gamma H=H$.
\end{proposition}
\begin{proof*}
	We split $\Hil$ as in \eqref{eq:H-split}, and have to construct a conjugation on each of the three summands. Since $H$ has no components in the first summand, and the second (complex linear) summand is contained in $H$, we can pick arbitrary conjugations on the first two summands. In other words, it is sufficient to consider the case where $H$ is standard.

	Recall that standard subspaces $H$ are in one-to-one correspondence with their Tomita operators $S_H$ via $H=\ker(1-S_H)$. Therefore a conjugation $\Gamma$ preserves a standard subspace $H$ if and only if it commutes with $S_H$ on the domain $H+iH$. Proceeding to the polar decomposition, this is also equivalent to $\Gamma$ commuting with both, the modular conjugation $J_H$, and the modular operator $\Delta_H$ (on its domain).
	
	We therefore need to construct a conjugation $\Gamma$ commuting with both modular data, $J$ and $\Delta$, of $H$. (For brevity, we drop the index ``$H$'' on these operators during this proof.) The proof is based on the relation
	\begin{align}\label{eq:JDelta-Exchange}
		J\Delta J=\Delta^{-1}\,.
	\end{align}
	Let $\Hil_1\subset\Hil$ denote the spectral subspace of spectrum of $\Delta$ in $\{1\}$ (which is zero in the factor situation $H\cap \accentset{\circ}{H}=\{0\}$). Furthermore, split $\Hil\ominus\Hil_1$ into subspaces $\Hil_<$ and $\Hil_>$, corresponding to spectrum of $\Delta$ in $[0,1]$ and $[1,\infty)$, respectively. We then have $\Hil=\Hil_<\oplus\Hil_1\oplus\Hil_>$, and $\Delta$ leaves all three subspaces invariant. In view of \eqref{eq:JDelta-Exchange}, we see that $J\Hil_<=\Hil_>$, $J\Hil_1=\Hil_1$, and $J\Hil_>=\Hil_<$.
	
	The modular operator $\Delta$ restricts to a (complex linear, bounded) self-adjoint operator on~$\Hil_<$, and thus we find a conjugation $\Gamma_<$ on $\Hil_<$ that commutes with this restriction. Taking into account $J\Hil_>=\Hil_<$, the conjugation $J\Gamma_<J$ is seen to be well-defined on $\Hil_>$, and leave this space invariant. Furthermore, $J\Gamma_<J$ commutes with the restriction of $\Delta$ to $\Hil_>$ because of \eqref{eq:JDelta-Exchange}.
	
	Finally, on $\Hil_1$, the modular operator restricts to the identity. Hence $J|_{\Hil_1}$ is a conjugation on this space that commutes with $\Delta|_{\Hil_1}$. Summarizing this discussion,
	\begin{align}
		\Gamma
		:=
		\Gamma_<\oplus J| _{\Hil_1}\oplus (J\Gamma_<J)|_{\Hil_>}
	\end{align}
	is a conjugation on $\Hil$ that commutes with $\Delta$. By construction, it also commutes with $J$. Thus $\Gamma H=H$.	
\end{proof*}


\medskip

It has to be mentioned, however, that for a general inclusion $\Hti\subset H$ of standard subspaces, a conjugation preserving both $H$ and $\Hti$ need not exist. Our counterexample is that of a half-sided modular inclusion, i.e. an inclusion $\Hti\subset H$ satisfying $\Delta_H^{-it}\Hti\subset \Hti$ for $t\geq0$.

\begin{lemma}
     Let $\Hti\subset H$ be a non-trivial half-sided modular inclusion of standard subspaces. Then there exists no conjugation $\Gamma$ with $\Gamma\Hti=\Hti$ and $\Gamma H=H$.
\end{lemma}
\begin{proof*}
	Suppose $\Gamma$ is a conjugation with $\Gamma H=H$ and $\Gamma\Hti=\Hti$. Then $\Gamma$ commutes with both modular operators, $\Delta_H$ and $\Delta_{\Hti}$. In view of the anti-linearity of $\Gamma$ and $\Gamma^2=1$, this implies in particular that $\Gamma\Delta_H^{-it}\Delta_{\Hti}^{it}\Gamma=\Delta_H^{it}\Delta_{\Hti}^{-it}$ for all $t\in\Rl$.

	But for a half-sided modular inclusion, there exists a unitary one-parameter group $T$ with positive generator such that $\Hti=T(1)H$ and
	\begin{align}\label{eq:characteristic-translation-function}
		\Delta_H^{-it}\Delta_{\Hti}^{it}=T(e^{2\pi t}-1)\,.
	\end{align}
	See \cite{Longo:2008} for a proof of these facts in the standard subspace setting, and \cite{ArakiZsido:2005,Wiesbrock:1993-1} for the original von Neumann algebraic situation.

    Setting $x(t):=e^{2\pi t}-1$, we therefore find from \eqref{eq:characteristic-translation-function}
	\begin{align}\label{eq:TG}
		\Gamma T(x(t))\Gamma
		=
		\Gamma \Delta_H^{-it}\Delta_{\Hti}^{it}\Gamma
		=
		\Delta_H^{it}\Delta_{\Hti}^{-it}
		=
		T(e^{-2\pi t}-1)
		=
		T(-\tfrac{x(t)}{1+x(t)})\,,
	\end{align}
	i.e. $\Gamma T(x)\Gamma=T(-\tfrac{x}{1+x})$ for all $x>-1$. But this leads to a contradiction: Since $T$ is a one-parameter group, and $\Gamma^2=1$, we have, $x>-\frac{1}{2}$,
	\begin{align*}
		\Gamma T(2x)\Gamma
		=
		\Gamma T(x)\Gamma\cdot\Gamma T(x)\Gamma
		=
		T(-\tfrac{x}{1+x})\cdot T(-\tfrac{x}{1+x})=
		T(-\tfrac{2x}{1+x})\,.
	\end{align*}
	But on the other hand, using \eqref{eq:TG}, we also have
	\begin{align*}
		\Gamma T(2x)\Gamma
		=
		T(-\tfrac{2x}{1+2x})\,.
	\end{align*}
	Hence $T(-\tfrac{2x}{1+x})=T(-\tfrac{2x}{1+2x})$ for all $x>-\frac{1}{2}$, which is possible if and only if $T(x)=1$ for all $x\in\Rl$. Thus $\Hti=T(1)H=H$, which contradicts the non-triviality $\Hti\neq H$ of the inclusion.
\end{proof*}

\medskip

We leave it as an open problem to characterize inclusions $\Hti\subset H$ of standard subspaces which allow for a conjugation preserving both $H$ and $\Hti$.

\subsection{$\ell^p$-conditions and Fermionic second quantization}\label{SSec_fermions}

To conclude this section, we also briefly consider the Fermionic case, where one again starts from a closed real subspace $H\subset\Hil$ as before, but proceeds to the Fermionic Fock space $\F_-(\Hil)$, and the von Neumann algebra $\M_-(H)$ generated by the Fermi field operators $\Phi[h]$, $h\in H$. The structure of the map $H\mapsto\M_-(H)$ is analogous to the one discussed in Lemma~\ref{Lem_Gamma_delta}, with the commutant replaced by a twisted commutant (see, for example, \cite[Thm.~55]{Derezinski:2004}). We will denote the analogue of $\Xi_{H,X_1}$ \eqref{eq:Xi-HX} in the Fermi case by $\Xi^-_{H,X_1}$, with identical assumptions on $X_1$ as in Section~\ref{SSec_secondq}, but now on $\F_-(\Hil)$ instead of $\F(\Hil)$.

There exists a result about second quantization of modular nuclearity conditions in the Fermionic case.

\begin{theorem}{\bf \cite{Lechner:2005}}\label{theorem:fermi-old}
     Let $H\subset\Hil$ be a closed real subspace and $X_1$ a selfadjoint operator on $\Hil$, satisfying the same assumptions as in Thm.~\ref{theorem:Buchholz-Jacobi}, with the exception of the norm bound $\|X_1E_\pm\|<1$, which is not assumed.

	Then $\Xi^-_{H,X_1}$ is nuclear, and its nuclearity index can be estimated as
	\begin{align}\label{eq:fermi-bound-old}
		\nu_1(\Xi^-_{H,X_1})
		\leq
		\exp\left(2\|X_1E_+\|_1+2\|X_1E_-\|_1\right)
		<
		\infty\,.
	\end{align}
\end{theorem}

The absence of the condition on the norm of $X_1E_\pm\in\B(\Hil)$, and the sharper bound on $\nu_1(\Xi^-_{H,X_1})$, are consequences of the Pauli principle \cite{Lechner:2005}.

In our more abstract setting, we find

\begin{theorem}\label{theorem:fermi-new}
     Let $H\subset\Hil$ be a closed real subspace.
     \begin{enumerate}
     	\item If $X_1E_H$ is $\ell^p_\Rl$  for some $0<p\leq1$, then $\Xi^-_{H,X_1}$ is $p$-nuclear, and $\ell^q$ for $q>p/(1-p)$.
     	\item If $\Xi^-_{H,X_1}$ is $\ell^p$ for some $p>0$, then $X_1E_H$ is $\ell^p_\Rl$.
     	\item $\Xi^-_{H,X_1}$ is $\ell^p$ for all $p>0$ if and only if $X_1E_H$ is $\ell^p_\Rl$ for all $p>0$.
     \end{enumerate}
\end{theorem}
\begin{proof*}
	$a)$ 	The proof is similar to that of Thm.~\ref{theorem:second-quantization}, but simpler because we do not need to control the norm of $X_1E_H$. We again pick a conjugation $\Gamma$ on $\Hil$, and consider the doubled system $\underline{H}\subset\underline{\Hil}$, $\underline{X}$ \eqref{eq:doubledsystem}, invariant under $\underline{\Gamma}$. Then, as in the Bose case, we can write $\underline{H}=\underline{\Gamma}^+\K_++\underline{\Gamma}^-\K_-$ with complex closed subspaces $\K_\pm\subset\underline{\Hil}$, and the corresponding operators $X_1E_\pm$ lie in $\ell^p(\underline{\Hil})$ because $X_1E_H\in\ell^p_\Rl(\Hil)$. We again proceed to the joint upper bound $T\in\ell^p(\underline{\Hil})$, satisfying $T^2\geq| X_1E_\pm|^2$.
	
	Nuclearity estimates on
	\begin{align}
		\Xi_{\underline{H},\underline{X}_1}^-:\M_-(\underline{H})\to\F_-(\underline{\Hil})
		\,,\qquad
		\underline{A}\mapsto \underline{X}\underline{A}\,\underline{\Om}
	\end{align}
	are obtained by following \cite{Lechner:2005}: Denoting the eigenvalues of $T$ by $t_j$, it is shown there (p.~3051) that the corresponding Fermi second quantized orthonormal basis $\{\bxi_\bmu\}_\bmu$ of $\F_-(\underline{\Hil})$ satisfies
	\begin{align}
		\sum_\bmu|\langle\bxi_\bmu,\,\Xi_{\underline{H},\underline{X}_1}^-(\underline{A})\rangle|
		\leq
		\|\underline{A}\|\prod_{j=1}^\infty(1+2t_j)
		\,.
	\end{align}
	Note that the last expression is dominated by $\|\underline{A}\|e^{2\|T\|_1}$, and thus finite because $T$ is an element of $\ell^p$, $p\leq1$, which is contained in the trace class.
	
	If one estimates the $p$-th powers of the expectation values instead, one gets in a similar manner
		\begin{align}
		\sum_\bmu|\langle\bxi_\bmu,\,\Xi_{\underline{H},\underline{X}_1}^-(\underline{A})\rangle|^p
		&\leq
		\|\underline{A}\|^p\prod_{j=1}^\infty(1+(2t_j)^p)
		\,,
	\end{align}
	from which we read off
	\begin{align}
		\nu_p(\Xi_{\underline{H},\underline{X}_1}^-)
		\leq
		\left(
		\prod_{j=1}^\infty(1+(2t_j)^p)
		\right)^{1/p}
		\leq
		\exp\frac{\|(2T)^p\|_1}{p}
		=
		\exp\frac{\|2T\|_p^p}{p}
		<
		\infty\,.
	\end{align}
	Hence $\Xi_{\underline{H},\underline{X}_1}$ is $p$-nuclear, and by Lemma~\ref{lemma:p-nuc-and-ell-p-2}, also $\ell^q$ for all $q>p/(1-p)$.
	
	The conversion of these estimates to corresponding ones for the system without the doubling now follows as in the Bose case.
	
	$b)$ The Fermi field operator $\Phi[h]$, $h\in H$ (sum of Fermionic creation and annihilation operator), is bounded with norm $\|\Phi[h]\|\leq2\|h\|$, and an element of $\M_-(H)$ satisfying $\Phi[h]\Om=h$. Thus the composition of bounded and $\ell^p$ maps (where $P_1$ denotes the projection $\F_-(\Hil)\to\Hil$)
	\begin{align*}
		H&\xrightarrow{\Phi}\M_-(H)\xrightarrow{\Xi^-_{H,X_1}}\F_-(\Hil)\xrightarrow{P_1}\Hil
		\\
		h&\longmapsto\Phi[h]\longmapsto X\Phi[h]\Om=X_1h	
	\end{align*}
	is $\ell^p$, meaning that $h\mapsto X_1h$ is in $\ell^p(H,\Hil)$. But this is equivalent to $X_1E_H\in\ell^p$.
	
	$c)$ This is a direct consequence of $a)$ and $b)$.
\end{proof*}

\medskip

Also in the Fermi case, one can apply this general result to more concrete modular or energy nuclearity conditions. We refrain from giving the details here.

\section{$\ell^p$-properties of Laplace-Beltrami operators}\label{Sec_geo}

In our analysis of the modular $\ell^p$-condition for a free scalar field in Section~\ref{Sec_freefield}, we will consider the ground state of a free scalar field in a standard ultra-static spacetime, which is intimately related to the geometry of the Cauchy surface.
This means in particular that the modular $\ell^p$-condition is encoded in certain results in Riemannian geometry. Because these results may be of independent interest, we will now present them in a general context.

Let $(\mathcal{C},h)$ be a Riemannian manifold with metric $h_{ij}$ and let $A:=-\Delta+m^2$ be the modified Laplace operator for some fixed mass parameter $m>0$, where $\Delta$ is the Laplace-Beltrami operator. We will view $A$ as an operator in the Hilbert space $L^2(\mathcal{C})$, defined on the dense domain $C_0^{\infty}(\mathcal{C})$. This section collects a number of results on $A$, including in particular some long range estimates on the integral kernels of powers of $A$.

It is clear from a partial integration that $A$ is positive (and hence symmetric). When $\mathcal{C}$ is complete we additionally have the following result by Chernoff \cite{Chernoff:1973}:
\begin{theorem}\label{Thm_A_esa}
For any $n\in\mathbb{N}$, $A^n$ is essentially self-adjoint on $C_0^{\infty}(\mathcal{C})$ in $L^2(\mathcal{C})$.
\end{theorem}
For ease of notation we will use the symbol $A$ from now on to denote the unique self-adjoint extension $\overline{A}$.

For any $\alpha\in\mathbb{R}$ and $f\in C_0^{\infty}(\mathcal{C})$, the vector $A^{\alpha}f$ is in the domain of all powers of $A$, so it is a smooth function (cf.~e.g.~Corollary 6.4.9 in \cite{Hoermander:1997} and note that it suffices to show smoothness locally). When $\alpha\le 0$, then $A^{\alpha}$ is bounded, because $A\ge m^2>0$. Moreover, for every $\alpha\in\mathbb{R}$ the operator $A^{\alpha}$ defines an integral kernel $K_{\alpha}\in\mathcal{D}'(\mathcal{C}^{\times 2})$ by
\[
K_{\alpha}(f_1,f_2):= \langle f_1,A^{\alpha}f_2\rangle
\]
(cf.~e.g.~Theorem A.1 of \cite{Sanders:2013}.) For $n\in\mathbb{N}_0$, the integral kernel $K_n$ is supported on the diagonal. For negative integer powers we have the following regularity result:
\begin{theorem}\label{Thm_A-n}
Let $d$ be the dimension of $\mathcal{C}$ and $k,l,n\in\mathbb{N}_0$.
\begin{enumerate}
\item If $n>\frac34d+\frac12(k+1)$, then $K_{-n}\in C^k(\mathcal{C}^{\times 2})$.
\item If $n>\frac34d+k+l+\frac12$ and $\chi_1,\chi_2\in C_0^{\infty}(\mathcal{C})$, then the operator $A^k\chi_1A^{-n}\chi_2A^l$ is Hilbert-Schmidt.
\end{enumerate}
\end{theorem}
\begin{proof*}
Let us write $A_x$, resp.~$A_y$, for the differential operator $A$ acting on the variables $x$, resp.~$y$, and note that $(A_x+A_y)^nK_{-n}(x,y)=2^n\delta(x,y)$. Because $A$ is elliptic on $\mathcal{C}$, $A_x+A_y$ is elliptic on $\mathcal{C}^{\times 2}$ and we may use the calculus of Sobolev wave front sets \cite{Hoermander:1997} to see that $WF^{(s)}(K_{-n})=WF^{(s-2n)}(\delta)$. The right-hand side is empty when $s-2n<-\frac{d}{2}$, so the left-hand side is empty when $s<2n-\frac{d}{2}$. When $\alpha$ is a multiindex in $x$ and $y$ with $|\alpha|\le k$, then we may choose $s=k+d+1$ to find $WF^{(s-k)}(\partial^{\alpha}K_{-n})\subset WF^{(s)}(K_{-n})=\emptyset$ when $n>\frac34d+\frac12(k+1)$. Note that $d$ is half the dimension of $\mathcal{C}^{\times 2}$, so $\partial^{\alpha}K_{-n}(x,y)$ is continuous by Sobolev's Lemma (Corollary 6.4.9 in \cite{Hoermander:1997}) and hence $K_{-n}(x,y)$ is in $C^k(\mathcal{C}^{\times 2})$. This proves the first item. For the second item we note that $\chi_1A^{-n}\chi_2$
is
in $C^{2(k+l)}(\mathcal{C}^{\times 2})$, so acting with the operators $A^k$ and $A^l$ we obtain an integral kernel $K$ for $A^k\chi_1A^{-n}\chi_2A^l$ which is still in $C^0(\mathcal{C}^{\times 2})$ and compactly supported. This means that $K\in L^2(\mathcal{C}^{\times 2})$ and hence the operator is Hilbert-Schmidt.
\end{proof*}

In addition to the regularity of $K_{-n}$ it will also be useful to investigate fall-off properties of the kernels $K_{\alpha}$ for general $\alpha$. A fundamental result in this direction is
\begin{proposition}\label{Prop_Abound}
Let $\alpha\in\mathbb{R}$ and let $\chi,\tilde{\chi}\in C^{\infty}(\mathcal{C})$ such that $\chi$ is bounded, $\tilde{\chi}$ has compact support and the supports of $\chi$ and $\tilde{\chi}$ are separated by a distance $\delta>0$. Then we have in $\B(L^2({\cal C}))$ the bound
\[
\|\chi A^{\alpha}\tilde{\chi}\|<C(\alpha)\|\chi\|_{\infty}\|\tilde{\chi}\|_{\infty} m^{-\frac32}\delta^{-\alpha-1}
\left(1+\frac{|\alpha(\alpha+1)|}{2m\delta}\right)e^{-m\delta}
\]
for some $C(\alpha)>0$, which is independent of $\chi,\tilde{\chi}$ and $\delta$.
\end{proposition}
\begin{proof*}
The smooth function $F(\lambda):=(\lambda^2+m^2)^{\alpha}$ defines a tempered distribution with Fourier transform $\hat{F}(s)$. On $s>0$ the distribution $g(s):=(s/m)^{\nu}\hat{F}(s/m)$ with $\nu:=\alpha+\frac12$ satisfies $g''(s)+s^{-1}g'(s)-(1+\nu^2 s^{-2})g(s)=0$, which is a modified Bessel equation. This means that $g(s)=C_1K_{\nu}(s)$ for some constant $C_1$, where the modified Bessel function $K_{\nu}(s)$ satisfies $K_{\nu}(s)\le\sqrt{\frac{\pi}{2s}}e^{-s}\left(1+\frac{|4\nu^2-1|}{8s}\right)$ (\cite{GradshteynRyzhikJeffrey:2000} 8.451)
and hence
\[
\int_{\delta}^{\infty}ds\, |\hat{F}(s)|\le |C_1|\sqrt{\frac{\pi}{2}}m^{-\frac32}\delta^{-\alpha-1}
\left(1+\frac{|\alpha(\alpha+1)|}{2m\delta}\right)e^{-m\delta}.
\]
Note that the operator $\chi A^{\alpha}\tilde{\chi}$ is well-defined on $C_0^{\infty}(\mathcal{C})$. Following Proposition 1.1 of \cite{CheegerGromovTaylor:1982} we now exploit the unit propagation speed of the wave operator $\partial_s^2-\Delta$, together with the fact that $F$ and $\hat{F}$ are real and even. For any $f\in C_0^{\infty}(\mathcal{C})$ this yields
\begin{eqnarray}
\|\chi A^{\alpha}\tilde{\chi} f\|&=&\left\|\frac{1}{2\pi}\int ds\ \hat{F}(s)\chi\cos(s\sqrt{-\Delta})\tilde{\chi}f\right\|\nonumber\\
&=&\left\|\frac{1}{\pi}\int_{\delta}^{\infty}ds\ \hat{F}(s)\chi\cos(s\sqrt{-\Delta})\tilde{\chi}f\right\|\nonumber\\
&\le&\frac{1}{\pi}\int_{\delta}^{\infty}ds\ |\hat{F}(s)|\ \|\chi\|_{\infty}\|\tilde{\chi}\|_{\infty}\|f\|,\nonumber
\end{eqnarray}
where $\|\chi\|_{\infty}$ is the operator norm of multiplication by $\chi$ and similarly for $\tilde{\chi}$. Combining the two estimates yields the result.
\end{proof*}

The following lemma holds quite generally, even when $\mathcal{C}$ is not complete:
\begin{lemma}\label{Lem_PDObound}
Let $V\subset\mathcal{C}$ be an open region and let $\Psi$ be a partial differential operator of order $r$ with smooth coefficients supported in $V$. For any $R\in\mathbb{N}_0$ such that $2R\ge r$, there are $\eta_1,\ldots,\eta_R\in C^{\infty}(\mathcal{C})$ supported in $V$ such that $\|\Psi f\|^2\le \sum_{k=0}^R\|\eta_kA^k f\|^2$ for all $f\in C_0^{\infty}(\mathcal{C})$.
\end{lemma}
If we did not care about the supports of the $\eta_i$, but the derivatives of the coefficients of $\Psi$ were suitably bounded (e.g.~when they are compactly supported), then we might replace the $\eta_i$ by a single constant $C>0$. Using the fact that $A^{k+l}\ge m^{2l}A^k$ we would then find the basic estimate $\|\Psi f\|\le C\|A^Rf\|$ for some $C>0$. However, the point of the lemma is that for $\Psi$ with coefficients supported in $V$, one can find $\eta_k$ supported in the same set $V$ and satisfying the desired estimate.
\begin{proof*}
By taking complex conjugates of the coefficients of $\Psi$ we obtain another partial differential operator $\overline{\Psi}$. Note that $\|\Psi f\|^2\le\langle f,Xf\rangle$, where $X=\Psi^*\Psi+\overline{\Psi}^*\overline{\Psi}$ is a symmetric partial differential operator of even order $2r$ with real coefficients supported in $V$. We will show by induction over $r$ that for any such operator $X$ we can estimate
\[
|\langle f,Xf\rangle|\le \sum_{k=0}^R\|\eta_kA^kf\|^2
\]
for some $\eta_1,\ldots,\eta_R\in C^{\infty}(\mathcal{C})$, which proves the claim. We will make repeated use of the following observation: when $X_1$ and $X_2$ can be estimated in this way, then so can $X_1+X_2$, because for any $\eta_{1,k}$ and $\eta_{2,k}$ with support in $V$ we can find $\eta_k$ with support in $V$ such that $|\eta_k|^2\ge|\eta_{1,k}|^2+|\eta_{2,k}|^2$.
In addition we will make use of the fact that $X$ may be written as $X=\sum_{k=0}^rX_{2k}$ with
\[
X_{2k}=\nabla_{a_1}\cdots\nabla_{a_k}\xi_{2k}^{a_1\cdots a_{2k}}\nabla_{a_{k+1}}\cdots\nabla_{a_{2k}}
\]
for some smooth coefficient tensor fields $\xi_{2k}$ supported in $V$. (This may also be shown by induction over $r$.)

In the case $r=0$, $X$ is simply a multiplication operator by a real function $\xi_0$ supported in $V$, and it suffices to choose $\eta_0$ such that $|\eta_0|^2\ge |\xi_0|$. Now suppose that the claim holds for all symmetric operators of order $0,\ldots,2(r-1)$. By our observation and the induction hypothesis it suffices to estimate the operator $X_{2r}$. For this we use the fact that the metric $h_{ij}$ defines an inner product on each tangent and cotangent space, which can be extended to the tensor bundle of each type $(k,l)$. We denote the corresponding norm by $\|.\|_h$ and by a partial integration and Cauchy's inequality we have
\begin{eqnarray}
|\langle f,X_{2r}f\rangle|&\le&\int_M\|\xi_{2r}\|_h\ \|\nabla_{a_1}\cdots\nabla_{a_k}f\|_h^2\nonumber\\
&\le&\int_M|\eta|^2\|\nabla_{a_1}\cdots\nabla_{a_k}f\|_h^2=\langle f,Yf\rangle,\nonumber
\end{eqnarray}
where $\eta\in C^{\infty}(\mathcal{C})$ is supported in $V$ and satisfies $|\eta|^2\ge\|\xi_{2r}\|_h$ and $Y$ is defined by
\[
Y:=(-1)^r\nabla^{a_r}\cdots\nabla^{a_1}|\eta|^2\nabla_{a_1}\cdots\nabla_{a_r}.
\]
We now set
\[
Y_{2r}:=\left\{\begin{array}{ll}
A^{\frac{r}{2}}|\eta|^2 A^{\frac{r}{2}}&r\mathrm{\ is\ even}\\
-A^{\frac{r-1}{2}}\nabla^a|\eta|^2 \nabla_aA^{\frac{r-1}{2}}&r\mathrm{\ is\ odd}
\end{array}\right.
\]
and we note that $Y-Y_{2r}$ is symmetric with real coefficients supported in $V$ and of order $<2r$. We then have
\[
|\langle f,X_{2r}f\rangle|\le \langle f,Yf\rangle\le
\langle f,Y_{2r}f\rangle+|\langle f,(Y-Y_{2r})f\rangle|,
\]
where the second term can again be estimated by the induction hypothesis. It only remains to prove the estimate for $Y_{2r}$. However, for even $r$ the estimate is immediate, and when $r$ is odd it follows from the fact that
\begin{eqnarray}
-\nabla^a|\eta|^2\nabla_a&=&A\frac{|\eta|^2}{2m^2}A-\Delta\frac{|\eta|^2}{2m^2}\Delta
+\frac12(\Delta|\eta|^2)-\frac{m^2}{2}|\eta|^2\nonumber\\
&\le&A\frac{|\eta|^2}{2m^2}A+\frac12(\Delta|\eta|^2)\nonumber
\end{eqnarray}
and hence
\[
-A^{\frac{r-1}{2}}\nabla^a|\eta|^2 \nabla_aA^{\frac{r-1}{2}}\le
A^{\frac{r+1}{2}}|\eta_{\frac{r+1}{2}}|^2A^{\frac{r+1}{2}}+
A^{\frac{r-1}{2}}|\eta_{\frac{r-1}{2}}|^2A^{\frac{r-1}{2}},
\]
where $\eta_{\frac{r+1}{2}}:=\frac{\eta}{\sqrt{2}m}$ and $|\eta_{\frac{r-1}{2}}|^2\ge \frac12(\Delta|\eta|^2)$. This completes the proof.
\end{proof*}

\begin{theorem}\label{Thm_Abound}
Let $\alpha,\beta,\gamma\in\mathbb{R}$ and let $\chi,\tilde{\chi}\in C_0^{\infty}(\mathcal{C})$ such that $\chi\equiv 1$ on a neighborhood of $\mathrm{supp}(\tilde{\chi})$. Then $A^{\beta}(1-\chi)A^{\alpha}\tilde{\chi}A^{\gamma}$ is bounded.
\end{theorem}
\begin{proof*}
When $\beta=\gamma=0$ the result follows immediately from Proposition \ref{Prop_Abound}. Now let $V,\tilde{V}\subset\mathcal{C}$ be open subsets such that $V^c:=\mathcal{C}\setminus\overline{V}$ contains the support of $1-\chi$, $\tilde{V}$ that of $\tilde{\chi}$, and $\overline{V^c}$ and $\overline{\tilde{V}}$ are disjoint. Note that $\tilde{V}$ is relatively compact, so that $\overline{V^c}$ and $\overline{\tilde{V}}$ are separated by a minimal distance $\delta>0$.

For any $c\in\mathbb{N}_0$, $A^c\overline{\tilde{\chi}}$ is a partial differential operator with smooth coefficients supported in $\tilde{V}$, so we may apply Lemma \ref{Lem_PDObound} to find $\tilde{\eta}_1,\ldots,\tilde{\eta}_c\in C_0^{\infty}(\tilde{V})$ such that
\[
\|A^c\overline{\tilde{\chi}}f\|^2\le\sum_{k=0}^c\|\tilde{\eta}_kA^kf\|^2
\]
for all $f\in C_0^{\infty}(\mathcal{C})$. Using Theorem \ref{Thm_A_esa} this result can be extended to all $f$ in the domain of $A^c$. Hence, for any $f\in C_0^{\infty}(\mathcal{C})$ and any $\theta\in C^{\infty}(\mathcal{C})$:
\[
\|A^c\overline{\tilde{\chi}}A^{\alpha}\overline{\theta}f\|^2\le\sum_{k=0}^c\|\tilde{\eta}_kA^{k+\alpha}\overline{\theta}f\|^2.
\]
If $\theta$ is bounded and supported in $V^c$ it follows again from Proposition \ref{Prop_Abound} that $\theta A^{k+\alpha}\overline{\tilde{\eta}_k}$ and hence also the adjoints $\tilde{\eta}_kA^{k+\alpha}\overline{\theta}$ are bounded. Therefore, $A^c\overline{\tilde{\chi}}A^{\alpha}\overline{\theta}$ and its adjoint $\theta A^{\alpha}\tilde{\chi}A^c$ are bounded too.

For any $b\in\mathbb{N}_0$ we now consider the partial differential operator
\[
\Psi:=A^b(1-\chi)-(1-\chi)A^b=\chi A^b-A^b\chi.
\]
We may choose a relatively compact open subset $U\subset\mathcal{C}$ which contains the support of $\chi(1-\chi)$ and such that $U\subset V^c$. Note that $\Psi$ has coefficients supported in $U$, so appealing again to Lemma \ref{Lem_PDObound} we may find $\eta_1,\ldots,\eta_b\in C_0^{\infty}(U)$ such that
\[
\|\Psi f\|^2\le\sum_{k=0}^b\|\eta_kA^k f\|^2
\]
for all $f\in C_0^{\infty}(\mathcal{C})$. By Theorem \ref{Thm_A_esa} this estimate can be extended to all $f$ in the domain of $A^b$, so for every $f\in C_0^{\infty}(\mathcal{C})$ we find
\[
\|\Psi A^{\alpha}\tilde{\chi}A^cf\|^2\le \sum_{k=0}^b\|\eta_kA^{k+\alpha}\tilde{\chi}A^cf\|^2.
\]
Now the operators $\eta_kA^{k+\alpha}\tilde{\chi}A^c$ and $(1-\chi)A^{b+\alpha}\tilde{\chi}A^c$ are bounded, by the second paragraph of this proof, and hence so are $\Psi A^{\alpha}\tilde{\chi}A^c$ and
\[
(1-\chi)A^{b+\alpha}\tilde{\chi}A^c+\Psi A^{\alpha}\tilde{\chi}A^c=A^b(1-\chi)A^{\alpha}\tilde{\chi}A^c.
\]

Finally, given any $\beta,\gamma\in\mathbb{R}$ we may choose $b,c\in\mathbb{N}_0$ such that $b\ge\beta$ and $c\ge\gamma$.
Then
\[
A^{\beta-b}(A^b(1-\chi)A^{\alpha}\tilde{\chi}A^c)A^{\gamma-c}=A^{\beta}(1-\chi)A^{\alpha}\tilde{\chi}A^{\gamma}
\]
is a product of three bounded operators, and hence bounded.
\end{proof*}

To improve on Theorem \ref{Thm_Abound} we may use the following lemma, which is adapted from \cite{Verch:1993_3}:
\begin{lemma}\label{Lem_lp}
Let $T$ be an operator in $L^2(\mathcal{C})$ defined on $C_0^{\infty}(\mathcal{C})$, let $\chi\in C_0^{\infty}(\mathcal{C})$ and assume that $T\chi A^n$ is bounded for all $n\in\mathbb{N}_0$. Then $T\chi A^{\beta}$ is $\ell^p$ for all $p>0$ and all $\beta\in\mathbb{R}$.
\end{lemma}
\begin{proof*}
Let $\beta\in\mathbb{R}$ and $N\in\mathbb{N}_0$ be arbitrary, set $\chi_1:=\chi$ and choose $\chi_2,\ldots,\chi_{2N+1}\in C_0^{\infty}(\mathcal{C})$ such that $\chi_{n+1}\equiv 1$ on $\mathrm{supp}(\chi_n)$ for $1\le n\le 2N$. We then have for all $1\le n\le 2N$ and $k,l\in\mathbb{N}_0$,
\[
\chi_nA^{k+l}=\chi_n\chi_{n+1}A^{k+l}=\chi_nA^k\chi_{n+1}A^l
\]
when acting on $C_0^{\infty}(\mathcal{C})$. Now pick $b,l\in\mathbb{N}_0$ such that $b>\beta$ and $l>\frac34d+\frac12$. Acting on the dense domain $A^{b-\beta}C_0^{\infty}(\mathcal{C})$ we then have:
\begin{eqnarray}
T\chi A^{\beta}&=&T\chi_1 A^b\chi_2\cdots\chi_{2N+1}A^{\beta-b}\nonumber\\
&=&T\chi_1A^b\left(\chi_2A^{Nl}A^{-Nl}\chi_3\right)\cdots\left(\chi_{2N}A^lA^{-l}\chi_{2N+1}\right)A^{\beta-b}\nonumber\\
&=&(T\chi_1A^{b+Nl})\left(\chi_2A^{-Nl}\chi_3A^{(N-1)l}\right)
\cdots\nonumber\\
&&\left(\chi_{2N-2}A^{-2l}\chi_{2N-1}A^l\right)\left(\chi_{2N}A^{-l}\chi_{2N+1}\right)A^{\beta-b}.\nonumber
\end{eqnarray}
The middle $N$ factors in brackets define Hilbert-Schmidt operators by Theorem \ref{Thm_A-n}, and the first and last factors are bounded (by assumption). Because Hilbert-Schmidt operators form an ideal in the bounded operators, the lemma follows.
\end{proof*}

Combining Lemma \ref{Lem_lp} with Theorem \ref{Thm_Abound} immediately yields
\begin{corollary}\label{Cor_Alp}.
Under the assumptions of Theorem \ref{Thm_Abound}, $A^{\beta}(1-\chi)A^{\alpha}\tilde{\chi}A^{\gamma}$ is $\ell^p$ for all $p>0$.
\end{corollary}


\section{The modular $\ell^p$-condition for free scalar fields}\label{Sec_freefield}

We now turn to the study of the modular $\ell^p$-condition for a real free scalar quantum field. On a general globally hyperbolic spacetime $M$ we assume that the field $\phi$ satisfies the Klein-Gordon equation
\[
\Box\phi(x)+\mathcal{V}(x)\phi(x)=0
\]
for some real valued smooth potential energy function $\mathcal{V}(x)$, where $\Box$ is the wave operator. It is well understood how to quantize this linear field equation in terms of the Weyl algebra $\mathcal{W}(M)$, how to describe quasi-free states and the Hadamard property, and how to obtain a net of sub-algebras $\mathcal{W}(O)$ corresponding to causally convex regions $O\subset M$ (see e.g.~\cite{BrunettiFredenhagenVerch:2001}). Using these notions we may formulate our main result.
\begin{theorem}\label{Thm_qfhad}
Every quasi-free Hadamard state on the Weyl algebra of a real free scalar quantum field on a globally hyperbolic spacetime satisfies the modular $\ell^p$-condition.
\end{theorem}

We do not need to assume that the field is generally covariant, so the potential $\mathcal{V}(x)$ may be quite arbitrary. Indeed, by Remark \ref{Rem_deformation} we can establish the modular $\ell^p$-condition by spacetime deformation, because this process preserves the quasi-free and Hadamard properties of states. During the deformation we may also deform the potential $\mathcal{V}(x)$ to the constant $m^2$ for some $m>0$, and we may deform the spacetime to an ultra-static one. In conclusion, Theorem \ref{Thm_deformation} and Remark \ref{Rem_deformation} reduce the problem to that of a massive, minimally coupled scalar field on an ultra-static spacetime and diamond regions based on regular Cauchy pairs in a fixed Cauchy surface.

A further reduction follows from Theorem \ref{theorem:second-quantization}, which reduces the problem to the one-particle level. In Subsection \ref{SSec_symplectic} we reformulate the one-particle modular $\ell^p$-condition in terms of properties of the symplectic form. In \ref{SSec_Hadamard} we then verify this property for quasi-free Hadamard states.

\subsection{The modular $\ell^p$-condition and the symplectic form}\label{SSec_symplectic}

In this section we consider a quasi-free state on a Weyl algebra and relate the modular operator arising from its GNS-representation to the symplectic form in the underlying abstract symplectic space. In the next section, we then consider explicit symplectic spaces of Cauchy data for the Klein-Gordon equation, and prove the modular $\ell^p$-condition for the theory of a free scalar field.

Let us fix a pre-symplectic space $(D_{\mathbb{R}},\sigma)$, i.e.~$D_{\mathbb{R}}$ is a real vector space and $\sigma$ a (possibly degenerate) anti-symmetric bilinear form. The Weyl algebra $\mathcal{W}(D_{\mathbb{R}},\sigma)$ is generated by the Weyl operators $W(f)$ with $f\in D_{\mathbb{R}}$ \cite{BinzHoneggerRieckers:2004}. We let $D$ be the complexification of $D_{\mathbb{R}}$ with complex conjugation $\Gamma$, and we extend $\sigma$ in a Hermitean way.

A quasi-free state on $\mathcal{W}(D_{\mathbb{R}},\sigma)$ is determined by a real (possibly semi-definite) inner product $\mu$ on $D_{\mathbb{R}}$ such that
\begin{equation}\label{Eqn_sigmabound}
\frac14|\sigma(f_1,f_2)|^2\le \mu(f_1,f_1)\mu(f_2,f_2).
\end{equation}
The corresponding state $\omega_{\mu}$ is uniquely determined by $\omega_{\mu}(W(f))=e^{-\frac12\mu(f,f)}$. We will fix a choice of $\mu$ and write $\mathcal{K}$ for the Hilbert space completion of $D$ in the unique Hermitean inner product that extends $\mu$ on $D_{\mathbb{R}}$. $\Gamma$ extends to an anti-unitary involution on $\mathcal{K}$ (denoted by the same symbol) and by Equation (\ref{Eqn_sigmabound}) there is a unique operator $\Sigma$ on $\mathcal{K}$ such that $\mu(f_1,\Sigma f_2)=\frac{i}{2}\sigma(f_1,f_2)$. We note that $\Sigma$ is self-adjoint, $\|\Sigma\|\le 1$ and $\Gamma\Sigma\Gamma=-\Sigma$.

The GNS-representation of $\omega_{\mu}$ consists of the Fock space $\mathcal{F}(\mathcal{H})$ over a one-particle Hilbert space $\mathcal{H}$ with Fock vector $\Omega_{\mu}$. The one-particle space can be constructed from $\mathcal{K}$ by dividing out $\mathrm{ker}(1+\Sigma)$, defining the inner product
\[
\langle F_1,F_2\rangle:=\omega_2(F_1,F_2):=\mu(f_1,(1+\Sigma)f_2)
\]
for the equivalence classes $F_i=[f_i]$, and taking the completion in this inner product. We let $\kappa:\mathcal{K}\to\mathcal{H}$ be the canonical map that arises out of this construction. There is a unitary map $U$ from $\mathcal{H}$ onto the subspace $\mathrm{ker}(1+\Sigma)^{\perp}$ in $\mathcal{K}$, defined by $U\kappa:=\sqrt{1+\Sigma}$. Note that $U\kappa$ is bounded and has dense range in $\mathrm{ker}(1+\Sigma)^{\perp}$.

Let $R$ be the orthogonal projection in $\mathcal{K}$ onto the kernel of $1-|\Sigma|$. It is known\footnote{Cf.~\cite{ArakiYamagami:1982} Theorem 3.12, and note that $\mathcal{H}$ as a real Hilbert space, with $\mathrm{Im}\langle,\rangle$ as symplectic form, can be complexified to a generalized Fock polarization. The algebra $\pi_{\omega_{\mu}}(\mathcal{W})'$ corresponds to the subspace $\overline{\kappa(D_{\mathbb{R}})}$, and the commutant corresponds to its symplectic complement.} that $\Omega_{\mu}$ is cyclic for the commutant $\pi_{\omega_{\mu}}(\mathcal{W})'$ if and only if $R=0$. In general, the commutant generates the subspace $\mathcal{F}(\mathcal{H})'=\mathcal{F}(\mathcal{H}')$ (cf.~Section \ref{SSec_Definition}), which is also a Fock space, but with $\mathcal{H}'\subset\mathcal{H}$ defined by
\[
\mathcal{H}':=\overline{\kappa((1-R)\mathcal{K})}.
\]
By the spectral calculus, $U\kappa$ decomposes as a direct sum map from $(1-R)\mathcal{K}\oplus R\mathcal{K}$ to $U\mathcal{H}'\oplus U(\mathcal{H}')^{\perp}$.

Recall from Lemma \ref{Lem_Gamma_delta} that the modular data $J,\Delta$ of the state $\omega_{\mu}$ on the algebra $\mathcal{W}(D_{\mathbb{R}},\sigma)$ are of second quantized form. More precisely, $H':=\overline{\kappa((1-R)D_{\mathbb{R}})}$ in $\mathcal{H}$ is a standard subspace of $\mathcal{H}'$, and the one-particle Tomita operator\footnote{In this section, we denote the one-particle projections of the modular data by $s,j,\delta$, and their second quantized version by $S,J,\Delta$.} $s$ is determined by
\begin{align}
	s\kappa(f)
	&=
	\begin{cases}
		\kappa(\Gamma f)\,,\qquad f\in (1-R)\mathcal{K},\\
		0\,,\qquad\qquad f\in R\mathcal{K}\,,		
	\end{cases}
\end{align}
which is a one-particle version of \eqref{eq:SQ}, as in \eqref{eq:generalized-tomita}.

In order to gain better control over the (usually unbounded) one-particle modular operator $\delta$ we use the following result.
\begin{proposition}\label{Prop_ddelta}
Let $d:=\frac{1-\Sigma}{1+\Sigma}(1-R)$ and let $h:\mathbb{R}_{\ge0}\to\mathbb{R}$ be any continuous function, so that $h(d)$ is a self-adjoint operator. Then $X\circ\kappa:=\kappa\circ h(d)$ defines an operator $X$ on a dense domain of $\mathcal{H}$, which is essentially self-adjoint with closure $h(\delta)$.
\end{proposition}
\begin{proof*}
Since $d$ vanishes on $R\mathcal{K}$ and $\delta$ on $(\mathcal{H}')^{\perp}$, it suffices to consider the summands $(1-R)\mathcal{K}$ and $\mathcal{H}'$. This is tantamount to the special case $R=0$, which we now consider. The operator $\sqrt{1+\Sigma}$ commutes with $h(d)$ and maps the domain of $h(d)$ into a core for $h(d)$, so $\overline{X}=U^*h(d)U$. It only remains to verify that $U^*dU=\delta$. For this we note that the range of $U\kappa$ is the domain of $d^{\frac12}$ and the range of $\kappa$ is a core for $\delta^{\frac12}$. Furthermore, using the one-particle Tomita operator $s$ we may compute for all $f_1,f_2\in K+iK$:
\begin{eqnarray}
\langle\delta^{\frac12}\kappa(f_1),\delta^{\frac12}\kappa(f_2)\rangle&=&\langle s\kappa(f_2),s\kappa(f_1)\rangle\nonumber\\
&=&\langle \kappa(\Gamma f_2),\kappa(\Gamma f_1)\rangle\nonumber\\
&=&\langle \Gamma f_2,(1+\Sigma)\Gamma f_1\rangle\nonumber\\
&=&\langle f_1,(1-\Sigma)f_2\rangle\nonumber\\
&=&\langle d^{\frac12}U\kappa(f_1),d^{\frac12}U\kappa(f_2)\rangle,\nonumber
\end{eqnarray}
This entails $\|\delta^{\frac12}f\|=\|U^*d^{\frac12}Uf\|$ on the domain of $d^{\frac12}$, which shows that $U^*d^{\frac12}U$ and $\delta^{\frac12}$ have equal domains and because they are both strictly positive they must be equal (\cite{ReedSimon:1972} Theorem VIII.15).
\end{proof*}

Now let $K:=\overline{D_{\mathbb{R}}}$. For a real subspace $\tilde{K}\subset K$ with $\tilde{H}:=\kappa(\tilde{K})$ we wish to determine whether $\delta^{\alpha}|_{\tilde{H}}$ is $\ell^p_{\mathbb{R}}$ for all $p>0$ and $\alpha\in(0,\frac12)$.
\begin{proposition}\label{Prop_deltamod}
Define $c(\alpha):=2^{2\alpha}$ for $0<\alpha<\frac14$ and $c(\alpha):=1$ else. In the notations above we then have for all $p>0$ and $\alpha\in\mathbb{R}$:
\begin{enumerate}
\item $\|\delta^{\alpha}|_{\tilde{H}}\|_{\mathbb{R},p}=\|\delta^{\frac12-\alpha}|_{\tilde{H}}\|_{\mathbb{R},p}$.
\item $\|(1-\Sigma^2)^{\alpha}|_{\tilde{K}+i\tilde{K}}\|_p\le c(\alpha)\|\delta^{\alpha}|_{\tilde{H}}\|_{\mathbb{R},p}$.
\item If $\alpha\le\frac14$, then $\|\delta^{\alpha}|_{\tilde{H}}\|_{\mathbb{R},p}\le 2^{-2\alpha}c(\alpha)\|(1-\Sigma^2)^{\alpha}|_{\tilde{K}+i\tilde{K}}\|_p$.
\end{enumerate}
Here we used the functional calculus with the convention $0^{\alpha}=0$, even for $\alpha\le0$.
\end{proposition}
\begin{proof*}
For any $\alpha\in\mathbb{R}$ and $f\in \tilde{K}$ in the domain of $d^{2\alpha}$ we have:
\begin{eqnarray}
\langle \kappa(f),\delta^{2\alpha}\kappa(f)\rangle&=&\langle f,(1+\Sigma)d^{2\alpha}f\rangle\nonumber\\
&=&\langle f,(1+\Sigma)^{1-2\alpha}(1-\Sigma)^{2\alpha}f\rangle\nonumber\\
&=&\langle f,(1-\Sigma)^{1-2\alpha}(1+\Sigma)^{2\alpha}f\rangle\nonumber\\
&=&\langle \kappa(f),\delta^{1-2\alpha}\kappa(f)\rangle,\nonumber
\end{eqnarray}
where we used $f=\Gamma f$ and the fact that the left-hand side is real in the third line. The first item now follows from Lemma \ref{Lem_lpbound}.

Averaging the equalities above we see that
\[
\|\delta^{\alpha}\kappa(f)\|^2=\frac12\|((1+\Sigma)^{1-4\alpha}+(1-\Sigma)^{1-4\alpha} )^{\frac12}(1-\Sigma^2)^{\alpha}f\|^2.
\]
Using the spectral calculus and $\|\Sigma\|\le 1$ we can estimate $(1+\Sigma)^{1-4\alpha}+(1-\Sigma)^{1-4\alpha}$ from below by $2^{1-4\alpha}$ when $0<\alpha<\frac14$ and by $2$ else, and from above by $2$ when $0\le\alpha\le\frac14$ and by $2^{1-4\alpha}$ when $\alpha<0$. The second and third items then follow from Lemma \ref{Lem_lpbound} and the following two remarks. The restriction of $\kappa$ to $\tilde{K}$ is isometric, because $\langle f,\Sigma f\rangle=0$ for $f\in\tilde{K}$. $(1-\Sigma^2)^{\alpha}$ commutes with the complex conjugation on $\mathcal{K}$, which entails that $\|(1-\Sigma^2)^{\alpha}|_{\tilde{K}+i\tilde{K}}\|_p=\|(1-\Sigma^2)^{\alpha}|_{\tilde{K}}\|_{\mathbb{R},p}$, by Remark \ref{Rem_lpR}.
\end{proof*}

To analyze the $\alpha$-dependence of the $\ell^p$ property we may use
\begin{lemma}\label{Lem_noalpha}
Let $X$ be a bounded positive operator on a complex Hilbert space $\mathcal{K}$, $E_0$ the projection onto its kernel and $P$ any other projection.
\begin{enumerate}
\item If $n\in\mathbb{N}$ and $\alpha\ge2^{-n}$, then $\|X^{\alpha}P\|_{2^np}\le\|X\|^{\alpha-2^{-n}}\|PXP\|_p^{2^{-n}}$.
\item If $X^{\alpha}P$ is $\ell^p$ for some $\alpha\le0$ and $p>0$, then $\|(1-E_0)P\|_p$ is finite too. In particular, if $E_0=0$, then $P$ has a finite dimensional range. (Here $X^{\alpha}$ is defined with the convention of Proposition \ref{Prop_deltamod}.)
\end{enumerate}
\end{lemma}
\begin{proof*}
By the polar decomposition, $\sqrt{X}P=U\sqrt{PXP}$ for some partial isometry $U$. Therefore,
$\|P\sqrt{X}P\|_{2p}\le\|\sqrt{X}P\|_{2p}=\|\sqrt{PXP}\|_{2p}=\|PXP\|_p^{\frac12}$. By induction we find $\|PX^{2^{-n}}P\|_{2^np}\le\|PXP\|_p^{2^{-n}}$ for all $n\in\mathbb{N}$, and hence
$\|X^{2^{-n}}P\|_{2^np}=\|PX^{2^{1-n}}P\|_{2^{n-1}p}^{\frac12}\le\|PXP\|_p^{2^{-n}}$. When $\alpha\ge 2^{-n}$ we may estimate $\|X^{\alpha}P\|_{2^np}\le\|X^{\alpha-2^{-n}}\|\cdot\|X^{2^{-n}}P\|_{2^np}\le\|X\|^{\alpha-2^{-n}}\|PXP\|_p^{2^{-n}}$.

On the other hand, if $X^{\alpha}P$ is $\ell^p$ for some $\alpha\le0$ and $p>0$, then $X^{-\alpha}$ is bounded and hence $X^{-\alpha}X^{\alpha}P=(1-E_0)P$ is $\ell^p$ too (recalling the convention of Proposition \ref{Prop_deltamod}). When $E_0=0$ this means that $P$ must have finite dimensional range.
\end{proof*}

\begin{corollary}\label{Cor_noalpha}
Let $P$ denote the orthogonal projection in $\K$ onto $\tilde{K}+i\tilde{K}$, and $p>0$.
\begin{enumerate}
\item $\|P(1-\Sigma^2)P\|_p^{\frac12}\le\|\delta^{\frac14}|_{\tilde{H}}\|_{\mathbb{R},2p}$.
\item If $\alpha\in(0,\frac14]$, then $\|\delta^{\frac12-\alpha}|_{\tilde{H}}\|_{\mathbb{R},p}=\|\delta^{\alpha}|_{\tilde{H}}\|_{\mathbb{R},p}\le2^{-2\alpha}c(\alpha)\|P(1-\Sigma^2)P\|_{2^{-n}p}^{2^{-n}}$ for all $n\in\mathbb{N}$ such that $\alpha\ge2^{-n}$.
\item If $\delta^{\alpha}|_{\tilde{H}}$ is $\ell^p_{\mathbb{R}}$ for some $\alpha\not\in(0,\frac12)$ and $p>0$, then $\|(1-R)P\|_p$ is finite. In particular, if $R=0$, then $\tilde{K}$ is finite dimensional.
\end{enumerate}
\end{corollary}
\begin{proof*}
This follows from Proposition \ref{Prop_deltamod} and Lemma \ref{Lem_noalpha} with $X=1-\Sigma^2$. For the first item we use the estimate
\[
\|P(1-\Sigma^2)P\|_p^{\frac12}=\|\sqrt{1-\Sigma^2}P\|_{2p}\le\|(1-\Sigma^2)^{\frac14}P\|_{2p}\le\|\delta^{\frac14}|_{\tilde{H}}\|_{\mathbb{R},2p}.
\]
For the second item we estimate
\[
\|\delta^{\alpha}|_{\tilde{H}}\|_{\mathbb{R},p}\le2^{-2\alpha}c(\alpha)\|(1-\Sigma^2)^{\alpha}P\|_p\le2^{-2\alpha}c(\alpha)
\|P(1-\Sigma^2)P\|_{2^{-n}p}^{2^{-n}}.
\]
For the third item we may assume $\alpha\le0$, so $\|(1-\Sigma^2)^{\alpha}P\|_p\le\|\delta^{\alpha}|_{\tilde{H}}\|_{\mathbb{R},p}$ is finite and the claim follows from the lemma.
\end{proof*}

In our application we will be interested in infinite dimensional spaces $\tilde{K}$ and typically $R=0$, so we cannot expect the one-particle modular $\ell^p$-condition to hold for $\alpha\not\in(0,\frac12)$ and any $p>0$. For the other values it is necessary and sufficient to show that $\sqrt{1-\Sigma^2}P$ is $\ell^p$ for all $p>0$.

\subsection{Modular $\ell^p$-condition for quasi-free Hadamard states}\label{SSec_Hadamard}

We will now establish the modular $\ell^p$-condition for quasi-free Hadamard states. Let us first review some facts and notations. A standard ultra-static globally hyperbolic spacetime is of the form $M=\mathbb{R}\times\mathcal{C}$ with metric $g=dt^2-h$, where the Killing time $t$ is given by projection onto the factor $\mathbb{R}$ and $h$ is a complete Riemannian metric on $\mathcal{C}$ which is independent of $t$ \cite{Kay:1978}. The Klein-Gordon equation then reduces to
\[
(\partial_t^2+A)\phi=0,
\]
where we recall from Section \ref{Sec_geo} that $A=-\Delta+m^2$. We will rely on the well-posedness of the Cauchy problem of this equation, and we may work with the Cauchy surface $\{0\}\times\mathcal{C}\simeq\mathcal{C}$ \cite{Sanchez:2005}. For each open region $V\subset\mathcal{C}$ we denote the space of complex initial data supported in $V$ by $\mathcal{D}(V):=C_0^{\infty}(V)\oplus C_0^{\infty}(V)$ and its subspace of real data by $\mathcal{D}_{\mathbb{R}}(V)$. We view these spaces as subspaces of $L^2(\mathcal{C})\oplus L^2(\mathcal{C})$, and we denote the canonical complex conjugation on the latter space by $\Gamma$. The spaces $\mathcal{D}_{\mathbb{R}}(V)$ are symplectic, with the symplectic form determined by the canonical commutation relations:
\[
\sigma((\varphi_1,\pi_1),(\varphi_2,\pi_2)):=\langle\varphi_1,\pi_2\rangle-\langle\pi_1,\varphi_2\rangle.
\]
The Weyl algebra $\mathcal{W}(M)$ is constructed from the symplectic space $\mathcal{D}_{\mathbb{R}}(\mathcal{C})$ and generated by the Weyl operators $W(f)$ with $f\in\mathcal{D}_{\mathbb{R}}(\mathcal{C})$. The sub-algebras $\mathcal{W}(D(V))$ for the diamond regions $D(V)$ are constructed analogously from $\mathcal{D}_{\mathbb{R}}(V)$.

A quasi-free state $\omega_{\mu}$ on $\mathcal{W}(M)$ is determined by a real inner product $\mu$ on $\mathcal{D}_{\mathbb{R}}(\mathcal{C})$, dominating the symplectic form as in (\ref{Eqn_sigmabound}). The two-point distribution $\omega_{\mu,2}$ can be split into symmetric and anti-symmetric parts $\omega_{\mu,\pm}(x,y):=\frac12\omega_{\mu,2}(x,y)\pm\frac12\omega_{\mu,2}(y,x)$, whose initial data are encoded by $\mu$ and $\frac{i}{2}\sigma$, respectively. By restriction, $\omega_{\mu}$ also determines a state on $\mathcal{W}(D(V))$ for any open $V\subset\mathcal{C}$. We will use subscripts $\mu,V$ for the various spaces and maps defined in Section \ref{SSec_symplectic}, to indicate their dependence on the choice of $\mu$ and $V$, and we will view $\mathcal{K}_{\mu,V}$ and $\mathcal{H}_{\mu,V}$ as subspaces of $\mathcal{K}_{\mu,\mathcal{C}}$ and $\mathcal{H}_{\mu,\mathcal{C}}$, respectively.

In a standard ultra-static spacetime, the massive free scalar field has a uniquely preferred ground state $\omega^0$, which is determined by the inner product
\[
\mu_0((\varphi_1,\pi_1),(\varphi_2,\pi_2)):=\frac12\langle\varphi_1,A^{\frac12}\varphi_2\rangle+\frac12\langle\pi_1,A^{-\frac12}\pi_2\rangle
\]
on $\mathcal{D}_{\mathbb{R}}(\mathcal{C})$ \cite{Kay:1978,Sanders:2013}. The operator $\Sigma:=\Sigma_{\mu_0,\mathcal{C}}$ can be written as the matrix
\begin{equation}\label{Eqn_Sigma0}
\Sigma=\left(\begin{array}{cc}
0&iA^{-\frac12}\\
-iA^{\frac12}&0
\end{array}\right)
\end{equation}
and for any open $V\subset\mathcal{C}$ we have $\Sigma_{\mu_0,V}=P_{\mu_0,V}\Sigma P_{\mu_0,V}$, where $P_{\mu_0,V}$ is the orthogonal projection in $\mathcal{K}_{\mu_0,\mathcal{C}}$ onto $\mathcal{K}_{\mu_0,V}$.
\begin{proposition}\label{Prop_sigma0nuclear}
For any regular Cauchy pair $(\tilde{V},V)$ in $\mathcal{C}$ and any $p>0$ the operator $P_{\mu_0,\tilde{V}}(1-\Sigma_{\mu_0,V}^2)P_{\mu_0,\tilde{V}}$ is $\ell^p$.
\end{proposition}
\begin{proof*}
We let $U:\mathcal{K}_{\mu_0}\to L^2(\mathcal{C})^{\oplus 2}$ be the unitary map defined by $U(\varphi,\pi):=\frac{1}{\sqrt{2}}(A^{\frac14}\varphi,A^{-\frac14}\pi)$ and we choose $\chi,\tilde{\chi}\in C_0^{\infty}(V)$ such that $\tilde{\chi}\equiv 1$ on $\tilde{V}$ and $\chi\equiv 1$ on a neighborhood of $\mathrm{supp}(\tilde{\chi})$. We then define operators $X$ and $\tilde{X}$ on $\mathcal{D}(\mathcal{C})$ by $\tilde{X}(\varphi,\pi):=(\tilde{\chi}\varphi,\tilde{\chi}\pi)$ and $X(\varphi,\pi):=(\chi\varphi,\chi\pi)$. These operators are closable and we denote their closures by the same symbol.
For any $f\in\mathcal{D}(\tilde{V})$ we have $\tilde{X}f=f$ and hence $P_{\mu_0,\tilde{V}}=\tilde{X}P_{\mu_0,\tilde{V}}$. Similarly, $P_{\mu_0,V}X=X$, which implies $1-P_{\mu_0,V}=(1-P_{\mu_0,V})(1-X)$. (Note how the ordering of these products matches the chosen support properties of $\tilde{\chi}$ and $\chi$.) Using $P_{\mu_0,\tilde{V}}P_{\mu_0,V}=P_{\mu_0,\tilde{V}}$ and $\Sigma^2=1$ \eqref{Eqn_Sigma0} we then have
\begin{eqnarray}
P_{\mu_0,\tilde{V}}(1-\Sigma_{\mu_0,V}^2)P_{\mu_0,\tilde{V}}&=&P_{\mu_0,\tilde{V}}\Sigma(I-P_{\mu_0,V})\Sigma P_{\mu_0,\tilde{V}}\nonumber\\
&=&|(1-P_{\mu_0,V})\Sigma P_{\mu_0,\tilde{V}}|^2\nonumber\\
&=&|(I-P_{\mu_0,V})(1-X)\Sigma\tilde{X}P_{\mu_0,\tilde{V}}|^2.\nonumber
\end{eqnarray}
Because the projections are bounded, it suffices to show that
\[
U(1-X)\Sigma \tilde{X}U^*=\frac{i}{2}\left(\begin{array}{cc}
0&A^{\frac14}(1-\chi)A^{-\frac12}\tilde{\chi}A^{\frac14}\\
-A^{-\frac14}(1-\chi)A^{\frac12}\tilde{\chi}A^{-\frac14}&0
\end{array}\right)
\]
is $\ell^p$ in $L^2(\mathcal{C})^{\oplus 2}$ for all $p>0$. This follows immediately from Corollary \ref{Cor_Alp}.
\end{proof*}

We wish to generalize Proposition \ref{Prop_sigma0nuclear} to more general quasi-free states $\omega_{\mu}$, i.e., we wish to determine whether
\begin{align*}
	P_{\mu,\tilde{V}}(1-\Sigma_{\mu,V}^2)P_{\mu,\tilde{V}}
	=
	P_{\mu,\tilde{V}}(1-\Sigma_{\mu,\mathcal{C}}^2)P_{\mu,\tilde{V}}
	+
	|(1-P_{\mu,V})\Sigma_{\mu,\mathcal{C}}P_{\mu,\tilde{V}}|^2
\end{align*}
is $\ell^p$ for all $p>0$ and all regular Cauchy pairs $(\tilde{V},V)$. Because both terms on the right-hand side are positive, both need to be $\ell^p$. Indeed, it is not hard to see that, in analogy to Corollary \ref{Cor_inclusions}, one may shrink the region $\tilde{V}$ and enlarge the region $V$ without spoiling this property of the symplectic form. In particular, the relative compactness of $V$ and non-emptiness of its complement in $\mathcal{C}$ are unnecessary.

For simplicity we will restrict attention to states $\omega_{\mu}$ which are locally quasi-equivalent to the ground state, i.e.~their restrictions to $\mathcal{W}(D(V))$ are quasi-equivalent to the restricted ground state for all relatively compact $V\subset\mathcal{C}$. It is known that this is the case if and only if
\[
\mu(f_1,f_2)=\mu_0(f_1,M_Vf_2)
\]
for all $f_1,f_2\in\mathcal{D}(V)$ for some bounded positive operator $M_V$ on $\mathcal{K}_{\mu_0,V}$ with bounded inverse such that $\sqrt{M_V+\Sigma_{\mu_0,V}}-\sqrt{1+\Sigma_{\mu_0,V}}$ is Hilbert-Schmidt \cite{ArakiYamagami:1982}.

\begin{lemma}\label{Lem_sigmanuclear}
Let $(\tilde{V},V)$ be a regular Cauchy pair in $\mathcal{C}$ and assume that $\omega_{\mu}$ and $\omega^0$ are quasi-equivalent states on $\mathcal{W}(D(V))$. Then, for any $p>0$:
\begin{align}
\|P_{\mu,\tilde{V}}(1-\Sigma_{\mu,V}^2)P_{\mu,\tilde{V}}\|_p\le\max\{4,2^{\frac{4}{p}-2}\}&\|M_V\|^{-1}\left(
(1+\|M_V\|^{-1})\|M_V-1\|_p+\right.\nonumber\\
&\left.\|P_{\mu_0,\tilde{V}}(1-\Sigma_{\mu_0,V}^2)P_{\mu_0,\tilde{V}}\|_p\right).\nonumber
\end{align}
\end{lemma}
\begin{proof*}
The map $U_Vf:=\sqrt{M_V}f$ defines a unitary isomorphism from $\mathcal{K}_{\mu,V}$ to $\mathcal{K}_{\mu_0,V}$ and we have $U_V\Sigma_{\mu,V}U_V^*=M_V^{-\frac12}\Sigma_{\mu_0,V}M_V^{-\frac12}$, because $\Sigma_{\mu_0,V}=M_V\Sigma_{\mu,V}$ on ${\cal D}(V)$. Moreover, for any $\tilde{V}\subset V$, $U_VP_{\mu,\tilde{V}}U_V^*$ is the orthogonal projection onto the range of $\sqrt{M_V}P_{\mu_0,\tilde V}$, which means in particular that
\[
M_V^{-\frac12}U_VP_{\mu,\tilde{V}}U_V^*=P_{\mu_0,\tilde{V}}M_V^{-\frac12}U_VP_{\mu,\tilde{V}}U_V^*.
\]
We may therefore rewrite $P_{\mu,\tilde{V}}(1-\Sigma_{\mu,V}^2)P_{\mu,\tilde{V}}$ as:
\begin{eqnarray}
&&P_{\mu,\tilde{V}}U_V^*M_V^{-\frac12}\left\{(M_V-1)+P_{\mu_0,\tilde{V}}(1-\Sigma_{\mu_0,V}^2)P_{\mu_0,\tilde{V}}+\right.\nonumber\\
&&\left.+\Sigma_{\mu_0,V}(1-M_V^{-1})\Sigma_{\mu_0,V}\right\}M_V^{-\frac12}U_VP_{\mu,\tilde{V}}.\nonumber
\end{eqnarray}
Here all operators are bounded, and the estimate follows from a repeated application of the quasi-norm inequality (\ref{Eqn_quasinorm}).
\end{proof*}

\begin{theorem}\label{Thm_LEQ}
For any relatively compact open region $V\subset\mathcal{C}$ and any quasi-free state $\omega_{\mu}$, the operator $M_V-1$ is $\ell^p$ for all $p>0$.
\end{theorem}
\begin{proof*}
This follows essentially from Proposition 3.8 of \cite{Verch:1994} and its proof, together with the following comments. \cite{Verch:1994} uses a spacetime formulation for the proof of its Proposition 3.8, but this is unitarily equivalent to the initial value formulation we use here. For any relatively compact region $W\subset\mathcal{C}$ containing $V$, the proof of Proposition 3.8 in \cite{Verch:1994} proves the existence of sequences of real elements $F_k,F'_k\in\mathcal{D}(W)$ such that
\[
\omega_{\mu,2}(F,F')-\omega^0_2(F,F')=\sum_{k=1}^{\infty}\sigma(F,F_k)\sigma(F',F'_k)=\sum_{k=1}^{\infty}\langle\Gamma F,\Sigma_{\mu}F_k\rangle\ \langle F'_k,\Sigma_{\mu}F'_k\rangle
\]
and such that
\[
\sum_{k=1}^{\infty}p(F_k)^{\lambda}p'(F'_k)^{\lambda}<\infty
\]
for all $\lambda>0$ and all continuous semi-norms $p,p'$ on $\mathcal{D}(W)$. (For the last property, see Appendix B loc.cit.) The $F_k$ and $F'_k$ correspond to sequences of vectors $\psi_k,\psi'_k\in\mathcal{K}_{\mu,W}$, and because $\omega_2$ is a distribution we find
\[
(M_V-1)\eta=\sum_{k=1}^{\infty}\langle\chi'_k,\eta\rangle\chi_k,
\]
where $\chi_k:=\Sigma_{\mu,W}\psi_k$ and $\chi'_k:=\Sigma_{\mu,W}\psi'_k$ and $\sum_{k=1}^{\infty}\|\chi_k\|^{\lambda}\|\chi'_k\|^{\lambda}<\infty$ for all $\lambda>0$. For the desired statement we merely need to project the vectors $\chi_k$ and $\chi'_k$ to the subspace $\mathcal{K}_{\mu,V}$. Because the projection does not increase lengths, this yields an $\ell^p$-representation of $M_V-1$ for all $p>0$.
\end{proof*}

We may now prove our main result of this section, Theorem \ref{Thm_qfhad}:
\begin{proof*}
We first consider a quasi-free Hadamard state $\omega_{\mu}$ in a standard ultra-static spacetime. From Theorem \ref{Thm_LEQ}, Lemma \ref{Lem_sigmanuclear} and Proposition \ref{Prop_sigma0nuclear} we see that $P_{\mu,\tilde{V}}(1-\Sigma_{\mu,V}^2)P_{\mu,\tilde{V}}$ is $\ell^p$ for all $p>0$ and all regular Cauchy pairs $(\tilde{V},V)$ in $\mathcal{C}$. By Proposition \ref{Prop_deltamod} and Corollary \ref{Cor_noalpha} this means that $\delta_{\mu,V}^{\alpha}|_{H_{\tilde{V}}}$ is $\ell^p$ for all $p>0$.

This shows that the first assumption of Thm.~\ref{thm:modular-nuclearity} is satisfied. Making use of Lemma~\ref{Lem_strictnorm}, the second assumption, the norm bound $\|\delta_{\mu,V}^\alpha|_{H_{\tilde{V}}}\|<1$, follows once we know $H_V\cap{\accentset{\circ}{H}}_V=\{0\}$, which means
that the local von Neumann algebra $\pi_{\omega_{\mu}}(\mathcal{W}(D(V)))''$ is a factor. Because all quasi-free Hadamard states are locally quasi-equivalent \cite{Verch:1994}, it suffices to verify this factor property in the ground state representation. There we use the fact that $\Sigma=\Sigma_{\mu_0,\mathcal{C}}$ is invertible, as can be seen from Equation (\ref{Eqn_Sigma0}). It then follows that $\Sigma_{\mu_0,V}=P_{\mu_0,V}\Sigma P_{\mu_0,V}$ is invertible in $\mathcal{K}_{\mu_0,V}$, which means that the local von Neumann algebra $\pi_{\mu_0}(\mathcal{W}(D(V)))''$ is a factor (by Lemma 3.3 of \cite{Verch:1994}).

Hence the assumptions of Thm.~\ref{thm:modular-nuclearity} are satisfied, and we conclude that the map $\Xi^{(\alpha)}_{\tilde{V},V,\omega_\mu}$ is $\ell^p$ for all $p>0$. We have now proved the theorem for all quasi-free Hadamard states of a massive scalar field in standard ultra-static spacetimes. For quasi-free Hadamard states in general globally hyperbolic spacetimes we may use a spacetime deformation argument, as explained in Remark \ref{Rem_deformation}.
\end{proof*}

\subsection{On non-Hadamard states satisfying the modular $\ell^p$-condition}

We have shown that all quasi-free Hadamard states of a free scalar field satisfy the modular $\ell^p$-condition, and the same is true for (finite) convex combinations of such states, by Proposition \ref{Prop_convexity}. In this section we critically review the idea whether one may simply replace the Hadamard condition by the modular $\ell^p$-condition, as a selection criterion for physically relevant states. We first give a class of examples of non-Hadamard, but otherwise well-behaved, quasi-free states with the modular $\ell^p$-condition. Then we discuss the question whether the modular $\ell^p$-condition admits much less well-behaved states.

\begin{example}
Our example of a non-Hadamard state with the modular $\ell^p$-property is a quasi-free state $\omega_{\mu}$ for the massive free scalar field on a standard ultra-static spacetime. In fact, we will choose $\omega_{\mu}$ to be quasi-equivalent to the ground state $\omega^0$, given by $\mu_0$. To define $\mu$, we choose an arbitrary vector $\psi\in \mathcal{K}_{\mu_0,\mathcal{C}}$ which is given by real initial data on $\mathcal{C}$, $\Gamma\psi=\psi$. Let $P_{\psi}$ be the orthogonal projection onto the linear space spanned by $\psi$ and set $M_{\mathcal{C}}:=I+P_{\psi}$. We note that $\Gamma M_{\mathcal{C}}\Gamma=M_{\mathcal{C}}$ and we define $\mu(f_1,f_2):=\mu_0(f_1,M_{\mathcal{C}}f_2)$. This defines a real inner product on $\mathcal{D}(\mathcal{C})$ which satisfies $\mu(f,f)\geq\mu_0(f,f)$ and therefore the bound (\ref{Eqn_sigmabound}), so it defines a quasi-free state $\omega_{\mu}$.

Note that $M_{\mathcal{C}}$ is bounded with bounded inverse and $M_{\mathcal{C}}-1$ has rank one, so it is $\ell^p$ for all $p>0$. This means that $\omega_{\mu}$ is quasi-equivalent to $\omega^0$ \cite{ArakiYamagami:1982}, and the same argument applies when restricting both states to the algebra of any relatively compact double cone region $\mathcal{W}(D(V))$. Moreover, $\omega_{\mu}$ satisfies the modular $\ell^p$-condition, by similar arguments as in the proof of Theorem \ref{Thm_qfhad}. Nevertheless, the vector $\psi$ defines data on $\mathcal{C}$ which may not be smooth, so $\mu-\mu_0$ may not be a smooth bi-solution to the Klein-Gordon equation and hence $\omega_{\mu}$ may not be Hadamard.
\end{example}

It is clear that the example above can be extended to operators on $\mathcal{K}_{\mu_0,\mathcal{C}}$ of the form $M_{\mathcal{C}}-1$ which are positive, of finite rank and which commute with~$\Gamma$. This provides a large class of quasi-free states satisfying the modular $\ell^p$-condition without being Hadamard. Although they are not Hadamard, these states are well-behaved in many other respects. They are all locally quasi-equivalent to the ground state, and the loss of regularity is restricted by the fact that $\mathcal{K}_{\mu_0,\mathcal{C}}$ consists of initial data with a specific Sobolev regularity. One might expect these states to be adiabatic Hadamard states in the sense of \cite{JunkerSchrohe:2002}. This would imply further good behavior, such as the validity of quantum energy inequalities \cite{FewsterSmith:2008}.

It is not clear whether there exist states satisfying the modular $\ell^p$-condition but with much worse behavior. In particular, we do not know whether all quasi-free states satisfying our condition must be locally primary or locally quasi-equivalent to each other. Even when local quasi-equivalence holds, it is unclear whether the operators $M_V-1$ of Lemma \ref{Lem_sigmanuclear} must be $\ell^p$ for all $p>0$, because the Lemma only proves that this condition is sufficient to conclude the analog of Proposition \ref{Prop_sigma0nuclear}. Of course the situation becomes even far less clear when considering states which are not quasi-free, or more general theories than a free scalar field.

\section{Discussion}\label{Sec_Discussion}

The main conclusion that one may draw from our investigation is that the modular $\ell^p$-condition, which we introduced as an extension of the modular nuclearity condition known from Minkowski space, is an interesting additional tool in the study of generally covariant quantum field theories. It can be defined in a quite general setting and behaves well w.r.t.~general covariance, as we demonstrated in Section \ref{Sec_Modnuc}. Although the physical interpretation of the modular $\ell^p$-condition is not fully clear, it seems a reasonable condition to impose, because it holds for all quasi-free Hadamard states of free scalar fields in all globally hyperbolic space-times, as we established in Sections \ref{Sec_secondq}, \ref{Sec_geo}, and \ref{Sec_freefield}.

As mentioned in the Introduction, the modular $\ell^p$-condition implies the split property, which expresses the statistical independence of observables located in space-like separated regions. In the context of generally covariant theories, this property was recently discussed by Fewster \cite{Fewster:2015}. For a free scalar field, our results confirm several of his findings.

We expect that the modular $\ell^p$-condition can be extended to other systems, such as free Fermions (using Thm.~\ref{theorem:fermi-new}) and the Proca field. For systems with gauge symmetries, such as free electromagnetism, there might be obstructions. Due to Gauss' law, the local von Neumann algebras need not be factors in that case, so the proof of our Lemma \ref{Lem_strictnorm} no longer holds. Whether the required norm bound still holds remains to be investigated.

Whether it is possible to use the modular $\ell^p$-condition as a sole selection criterion to select the physically relevant states of a system is not yet clear. Even when considering quasi-free states of a massive free scalar field in a standard ultra-static spacetime, it is not clear whether the modular $\ell^p$-condition implies local quasi-equivalence to the vacuum state. However, if it should turn out that this is not the case, one may still try to combine the modular $\ell^p$-condition with a condition of local quasi-equivalence, in order to obtain a good selection criterion for generally covariant quantum field theories.

Looking further ahead one may then wonder what other nice results might follow from such a selection criterion. One line of thought is to investigate whether the criterion can be used in order to define local fields for general (possibly interacting) theories, along the lines of \cite{FredenhagenHertel:1981,Bostelmann:2005}. Given the present setup, it may even be possible to derive the existence of generally covariant quantum fields in the sense of \cite{BrunettiFredenhagenVerch:2001}, and to provide a link between the $C^*$-algebraic setting of that paper and to the approach of \cite{HollandsWald:2010}, which is based on the operator product expansion (cf.\ \cite{Bostelmann:2005-2}).

\section*{Acknowledgements}

We would like to thank the MSc.\ and PhD.\ students at the University of Leipzig for inviting us to their informal discussion meetings, where our project was first conceived, and we are grateful to Rainer Verch for encouraging comments during the initial stages of this project. One of us (KS) would like to thank Chris Fewster for a question concerning quantum energy inequalities, posed at the 14th Marcel Grossman Meeting in Rome, 2015.


\footnotesize
\begin{thebibliography}{10}

\bibitem{Araki:1977}
H.~Araki.
\newblock {Relative Entropy of States of von Neumann Algebras II}.
\newblock {\em Publ. RIMS, Kyoto University}, (13):173--192, 1977.

\bibitem{ArakiYamagami:1982}
H.~Araki and S.~Yamagami.
\newblock {On quasi-equivalence of quasifree states of the canonical
  commutation relations}.
\newblock {\em Publ. RIMS, Kyoto University}, 18(2):283--338, 1982.

\bibitem{ArakiZsido:2005}
H.~Araki and L.~Zsido.
\newblock {Extension of the structure theorem of {B}orchers and its application
  to half-sided modular inclusions}.
\newblock {\em Rev. Math. Phys.}, 17:491--543, 2005.

\bibitem{BinzHoneggerRieckers:2004}
E.~Binz, R.~Honegger, and A.~Rieckers.
\newblock {Construction and uniqueness of the {$C^*$}-Weyl algebra over a
  general pre-symplectic space}.
\newblock {\em J. Math. Phys.}, 45(7):2885--2907, 2004.

\bibitem{Borchers:1965}
{{{{{{{{{{{{{{{{{{{{{{{{{H.-J.}}}}}}}}}}}}}}}}}}}}}}}}} Borchers.
\newblock {Local Rings and the Connection of Spin with Statistics}.
\newblock {\em Comm. Math. Phys.}, 1:281--307, 1965.

\bibitem{Borchers:2000}
{{{{{{{{{{{{{{{{{{{{{{{{{H.-J.}}}}}}}}}}}}}}}}}}}}}}}}} Borchers.
\newblock {On revolutionizing quantum field theory with Tomita's modular
  theory}.
\newblock {\em J. Math. Phys.}, 41:3604--3673, 2000.

\bibitem{Bostelmann:2005-2}
H.~Bostelmann.
\newblock {Operator product expansions as a consequence of phase space
  properties}.
\newblock {\em J.Math.Phys.}, 46:082304, 2005.

\bibitem{Bostelmann:2005}
H.~Bostelmann.
\newblock {Phase space properties and the short distance structure in quantum
  field theory}.
\newblock {\em J. Math. Phys.}, 46:052301, 2005.

\bibitem{BratteliRobinson:1987}
O.~Bratteli and D.~W. Robinson.
\newblock {\em {Operator Algebras and Quantum Statistical Mechanics I}}.
\newblock Springer, 1987.

\bibitem{BratteliRobinson:1997}
O.~Bratteli and D.~W. Robinson.
\newblock {\em {Operator Algebras and Quantum Statistical Mechanics II}}.
\newblock Springer, 1997.

\bibitem{BrunettiFredenhagenVerch:2001}
R.~Brunetti, K.~Fredenhagen, and R.~Verch.
\newblock {The generally covariant locality principle -- A new paradigm for
  local quantum physics}.
\newblock {\em Comm. Math. Phys.}, 237:31--68, 2003.

\bibitem{BuchholzDAntoniLongo:1990}
D.~Buchholz, C.~D'Antoni, and R.~Longo.
\newblock {Nuclear Maps and Modular Structures 2: Applications to Quantum Field
  Theory}.
\newblock {\em Comm. Math. Phys.}, 129:115, 1990.

\bibitem{BuchholzDAntoniLongo:1990-1}
D.~Buchholz, C.~D'Antoni, and R.~Longo.
\newblock {Nuclear maps and modular structures. I. General properties.}
\newblock {\em J. Funct. Anal.}, 88:233--250, 1990.

\bibitem{BuchholzFredenhagen:1982}
D.~Buchholz and K.~Fredenhagen.
\newblock {Locality and the Structure of Particle States}.
\newblock {\em Comm. Math. Phys.}, 84:1, 1982.

\bibitem{BuchholzJacobi:1987}
D.~Buchholz and P.~Jacobi.
\newblock {On the Nuclearity Condition for Massless Fields}.
\newblock {\em Lett. Math. Phys.}, 13:313, 1987.

\bibitem{BuchholzJunglas:1989}
D.~Buchholz and P.~Junglas.
\newblock {On the Existence of Equilibrium States in Local Quantum Field
  Theory}.
\newblock {\em Comm. Math. Phys.}, 121:255--270, 1989.

\bibitem{BuchholzLechner:2004}
D.~Buchholz and G.~Lechner.
\newblock {Modular nuclearity and localization}.
\newblock {\em Annales Henri Poincar{\'e}}, 5:1065--1080, 2004.

\bibitem{BuchholzWichmann:1986}
D.~Buchholz and E.~H. Wichmann.
\newblock {Causal Independence and the Energy Level Density of States in Local
  Quantum Field Theory}.
\newblock {\em Comm. Math. Phys.}, 106:321, 1986.

\bibitem{CheegerGromovTaylor:1982}
J.~Cheeger, M.~Gromov, and M.~Taylor.
\newblock {Finite Propagation Speed, Kernel Estimates for Functions of the
  Laplace Operator, and the Geometry of Complete Riemannian Manifolds}.
\newblock {\em J. Diff. Geo.}, 17:15--53, 1982.

\bibitem{Chernoff:1973}
P.~R. Chernoff.
\newblock {Essential self-adjointness of powers of generators of hyperbolic
  equations}.
\newblock {\em J. Funct. Anal.}, 12:401--414, 1973.

\bibitem{Conway:1978}
J.B. Conway.
\newblock {\em {Functions of One Complex Variable}}.
\newblock Springer, New York, Heidelberg, Berlin, 2 edition, 1978.

\bibitem{DappiaggiHackSanders:2014}
C.~Dappiaggi, T.~Hack, and K.~Sanders.
\newblock {Electromagnetism, local covariance, the Aharonov-Bohm effect and
  Gauss' law}.
\newblock {\em Commun. Math. Phys.}, 328:625--667, 2014.

\bibitem{Derezinski:2004}
J.~Derezinski.
\newblock {Introduction to Representations of Canonical Commutation and
  Anticommutation Relations}.
\newblock {\em Lect. Notes Phys.}, 695:63--143, 2006.

\bibitem{Dimock:1980}
J.~Dimock.
\newblock {Algebras of Local Observables on a Manifold}.
\newblock {\em Comm. Math. Phys.}, 77:219--228, 1980.

\bibitem{DoplicherHaagRoberts:1969}
S.~Doplicher, R.~Haag, and J.~E. Roberts.
\newblock {Fields, observables and gauge transformations. I}.
\newblock {\em Comm. Math. Phys.}, 13:1--23, 1969.

\bibitem{EckmannOsterwalder:1973}
{{{{{{{{{{{{{{{{{{{{{{{{{J.-P.}}}}}}}}}}}}}}}}}}}}}}}}} Eckmann and
  K.~Osterwalder.
\newblock {An application of Tomita's theory of modular Hilbert algebras:
  Duality for free Bose fields}.
\newblock {\em J. Funct. Anal.}, 13(1):1--12, 1973.

\bibitem{Fewster:2015}
C.~J. Fewster.
\newblock {The split property for locally covariant quantum field theories in
  curved spacetime}.
\newblock {\em Lett. Math. Phys.}, 105:1633--1661, 2015.

\bibitem{FewsterOjimaPorrmann:2004}
C.~J. Fewster, I.~Ojima, and M.~Porrmann.
\newblock {p-Nuclearity in a New Perspective}.
\newblock {\em Lett. Math. Phys.}, 73:1--15, 2005.

\bibitem{FewsterSmith:2008}
C.~J. Fewster and C.~J. Smith.
\newblock {Absolute quantum energy inequalities in curved spacetime}.
\newblock {\em Annales Henri Poincare}, 9:425--455, 2008.

\bibitem{FewsterVerch:2012}
C.~J. Fewster and R.~Verch.
\newblock {Dynamical locality and covariance: What makes a physical theory the
  same in all spacetimes?}
\newblock {\em Annales Henri Poincar{\'e}}, 13:1613--1674, 2012.

\bibitem{FlorigSummers:1997}
M.~Florig and S.~J. Summers.
\newblock {On the statistical independence of algebras of observables}.
\newblock {\em J. Math. Phys.}, 38:1318--1328, 1997.

\bibitem{FredenhagenHertel:1981}
K.~Fredenhagen and J.~Hertel.
\newblock {Local algebras of observables and point-like localized fields}.
\newblock {\em Comm. Math. Phys.}, 80:555, 1981.

\bibitem{FullingNarcowichWald:1981}
S.~A. Fulling, F.~J. Narcowich, and R.~M. Wald.
\newblock {Singularity structure of the two-point function in quantum field
  theory in curved spacetime}.
\newblock {\em Ann. Phys.}, 136:243--272, 1981.

\bibitem{GradshteynRyzhikJeffrey:2000}
I.~S. Gradshteyn, I.~M. Ryzhik, and A.~Jeffrey, editors.
\newblock {\em {Table of Integrals, Series, and Products}}.
\newblock Academic Press.

\bibitem{Haag:1996}
R.~Haag.
\newblock {\em {Local Quantum Physics - Fields, Particles, Algebras}}.
\newblock Springer, second edition, 1996.

\bibitem{HaagHugenholtzWinnink:1967}
R.~Haag, N.~M. Hugenholtz, and M.~Winnink.
\newblock {On the Equilibrium States in Quantum Statistical Mechanics }.
\newblock {\em Comm. Math. Phys.}, 5:215--236, 1967.

\bibitem{Hansen:2013}
F.~Hansen.
\newblock {The fast track to L{\"o}wner's theorem}.
\newblock {\em Lin. Alg. Appl.}, 438:4557--4571, 2013.

\bibitem{HollandsWald:2010}
S.~Hollands and R.~M. Wald.
\newblock {Axiomatic quantum field theory in curved spacetime}.
\newblock {\em Commun.Math.Phys.}, 293:85--125, 2010.

\bibitem{HollandsWald:2015}
S.~Hollands and R.~M. Wald.
\newblock {Quantum fields in curved spacetime}.
\newblock {\em Phys. Rep.}, 574:1--35, 2015.

\bibitem{Hoermander:1997}
L.~H{\"o}rmander.
\newblock {\em {Lectures on nonlinear hyperbolic differential equations}}.
\newblock Springer, 1997.

\bibitem{Jkel:2004}
C.~D. J{\"a}kel.
\newblock {The relation between KMS-states for different temperatures}.
\newblock {\em Annales Henri Poincare}, 5:579--606, 2004.

\bibitem{JunkerSchrohe:2002}
W.~Junker and E.~Schrohe.
\newblock {Adiabatic vacuum states on general space-time manifolds: Definition,
  construction, and physical properties}.
\newblock {\em Annales Poincare Phys.Theor.}, 3:1113--1182, 2002.

\bibitem{KadisonRingrose:1986}
R.~V. Kadison and J.~R. Ringrose.
\newblock {\em {Fundamentals of the Theory of Operator Algebras II - Advanced
  Theory}}.
\newblock 1986.

\bibitem{Kay:1978}
B.~S. Kay.
\newblock {Linear spin-zero quantum fields in external gravitational and scalar
  fields -- I. A one particle structure for the stationary case}.
\newblock {\em Comm. Math. Phys.}, 62:55--70, 1978.

\bibitem{KayWald:1991}
B.~S. Kay and R.~M. Wald.
\newblock {Theorems on the Uniqueness and Thermal Properties of Stationary,
  Nonsingular, Quasifree States on Space-Times with a Bifurcate Killing
  Horizon}.
\newblock {\em Phys. Rept.}, 207:49--136, 1991.

\bibitem{Lechner:2005}
G.~Lechner.
\newblock {On the existence of local observables in theories with a factorizing
  S-matrix}.
\newblock {\em J. Phys.}, A38:3045--3056, 2005.

\bibitem{Lechner:2008}
G.~Lechner.
\newblock {Construction of Quantum Field Theories with Factorizing
  {S}-Matrices}.
\newblock {\em Comm. Math. Phys.}, 277:821--860, 2008.

\bibitem{LeylandsRobertsTestard:1978}
P.~Leylands, J.~E. Roberts, and D.~Testard.
\newblock {Duality for Quantum Free Fields}.
\newblock {\em Preprint}, 1978.

\bibitem{Longo:2008}
R.~Longo.
\newblock {Lectures on Conformal Nets - Part 1}.
\newblock In {\em {Von Neumann algebras in Sibiu}}, pages 33--91. Theta, 2008.

\bibitem{Pietsch:1972}
A.~Pietsch.
\newblock {\em {Nuclear Locally Convex Spaces}}.
\newblock Springer, 1972.

\bibitem{ReedSimon:1972}
M.~Reed and B.~Simon.
\newblock {\em {Methods of Modern Mathematical Physics I - Functional
  Analysis}}.
\newblock Academic Press, 1972.

\bibitem{ReehSchlieder:1961}
H.~Reeh and S.~Schlieder.
\newblock {Bemerkungen zur Unit{\"a}r{\"a}quivalenz von lorentzinvarianten
  Feldern}.
\newblock {\em Il Nuovo Cimento}, 22(5):1051--1068, 1961.

\bibitem{Sanchez:2005}
M.~Sanchez.
\newblock {On the Geometry of Static Spacetimes}.
\newblock {\em Nonlinear Analysis}, 63:e455--e463, 2005.

\bibitem{Sanders:2009-1}
K.~Sanders.
\newblock {On the Reeh-Schlieder Property in Curved Spacetime}.
\newblock {\em Comm. Math. Phys.}, 288:271--285, 2009.

\bibitem{Sanders:2010}
K.~Sanders.
\newblock {Equivalence of the (generalised) Hadamard and microlocal spectrum
  condition for (generalised) free fields in curved spacetime}.
\newblock {\em Commun. Math. Phys.}, 295:485--501, 2010.

\bibitem{Sanders:2013}
K.~Sanders.
\newblock {Thermal equilibrium states of a linear scalar quantum field in
  stationary spacetimes}.
\newblock {\em Int. J. Mod. Phys. A}, 28:1330010, 2013.

\bibitem{Verch:1993_3}
R.~Verch.
\newblock {Nuclearity, split property and duality for the Klein-Gordon field in
  curved space-time}.
\newblock {\em Lett. Math. Phys.}, 29:297--310, 1993.

\bibitem{Verch:1994}
R.~Verch.
\newblock {Local definiteness, primarity and quasiequivalence of quasifree
  Hadamard quantum states in curved space-time}.
\newblock {\em Comm. Math. Phys.}, 160:507--536, 1994.

\bibitem{Verch:1999}
R.~Verch.
\newblock {Wavefront sets in algebraic quantum field theory}.
\newblock {\em Commun. Math. Phys.}, 205:337--367, 1999.

\bibitem{Verch:2001}
R.~Verch.
\newblock {A spin-statistics theorem for quantum fields on curved spacetime
  manifolds in a generally covariant framework}.
\newblock {\em Comm. Math. Phys.}, 223:261--288, 2001.

\bibitem{Wald:1994}
R.~M. Wald.
\newblock {\em {Quantum field theory in curved space-time and black hole
  thermodynamics}}.
\newblock Chicago University Press, 1994.

\bibitem{Wiesbrock:1993-1}
H.W. Wiesbrock.
\newblock {Half sided modular inclusions of von Neumann algebras}.
\newblock {\em Comm. Math. Phys.}, 157:83--92, 1993.

\end{thebibliography}
\end{document}